%% file: preprint.tex
\documentclass[12pt]{article}
\usepackage{fullpage}
\usepackage{english, german}
\usepackage{amsmath}
\usepackage{epsfig}
\usepackage{here}
\usepackage{epic, eepic}
\usepackage{latexsym,amssymb}
\usepackage{hhline}
\newcommand{\iceskip}{\\ \vspace{0mm}\\}
\begin{document}
\title{Sungrazing comets: Properties of nuclei and
in-situ detectability of cometary ions at $1\,$AU}
\author{M. Iseli, M. K\"uppers, W. Benz, P. Bochsler \\
Physikalisches Institut, University of Bern\\
Sidlerstrasse 5, CH-3012 Bern, Switzerland}
\date{}
\maketitle

\begin{abstract}
A one dimensional sublimation model for cometary nuclei
is used to derive size limits for the nuclei
of sungrazing comets, and to estimate
oxygen ion fluxes at $1\,$AU from their evaporation.
Given that none of the $\approx 300$ sungrazers detected by the
SOlar and Heliospheric Observatory (SOHO) was observed after disappearing
behind the sun, and that small nuclei with a radius of $\approx 3.5\,$m
could be observed, it is assumed that all SOHO sungrazers were completely
destroyed. For the case that sublimation alone is sufficient for destruction,
the model yields an upper size limit as a function of nuclear density
$\varrho$, albedo $A$ and perihelion distance $q$. If the density of the
nuclei is that typical of porous ice ($600\,$kg\,m$^{-3}$), the maximum size is
$63\,$m. These results confirm similar model calculations
by Weissman (1983). An analytical expression is derived
that approximates the model results well. We discuss possible modifications of
our results by different disruption mechanisms. While disruption by
thermal stress does not change the upper size limits significantly, they may
be somewhat increased by tidal disruption (up to 100\,m for a density of 
$600\,$kg\,m$^{-3}$) dependent on the isotropy of the sublimation process and
the tensile strength of the comet. Implications for the Kreutz family of 
sungrazers are discussed.

Oxygen ions from the sublimation of sungrazing comets form a tail. Fluxes
from this tail are sufficiently high to be measured at $1\,$AU by particle
detectors on spacecraft, but the duration of a tail crossing is only about half
an hour. Therefore the probability of a spacecraft actually encountering a tail
of an evaporating sungrazer is only of the order of two percent per year.
\end{abstract}
\newpage

% *****************************************************************
\section{Introduction}
Cometary nuclei are difficult to study. They cannot be spatially resolved
from earth and whenever they come close to the sun they are surrounded by a
coma which is several magnitudes brighter than the nucleus. The nucleus of
comet P/Halley has been investigated {\bfseries in situ} during the Giotto
fly-by in 1986, allowing an accurate determination of its size, shape, and
albedo (Keller et al. 1987). However, the mass of P/Halley could not be 
determined with sufficient accuracy to derive meaningful constraints on the 
density of the comet (Peale 1989).

Other methods to estimate some properties of the cometary nucleus
like size and albedo are remote observations of
comets far from the sun (e.g. Hainaut et al. 1995) and
interpretation of properties of the coma in observations at high
spatial resolution (e.g. Weaver et al. 1997). These observations
do not reveal any information about the density, material strength or
internal structure of comets.

Additional properties of cometary nuclei are revealed indirectly when
comets are disrupted. Up to 1996, 33 split comets have been observed
(Sekanina 1997). Five of these comets are possibly split by tidal forces,
in the other cases the cause of the disruption is unknown. Therefore
the study of disruption physics and that of the physical
nature of the cometary nucleus from cometary splittings in general
is difficult.

The disruption of comet D/Shoemaker-Levy 9 during a close encounter with
Jupiter and its subsequent collision with the planet has gained much
attention. A large number of studies about the nature of the comet has
been published, including the determination of the structure of its nucleus
from tidal breakup (Asphaug and Benz (1996), see Noll et al. (1996) for an
overview), but some controversy remains (Sekanina et al. 1998).

In this paper, we discuss the destruction of members  of the Kreutz group
of sungrazing comets, focussing on the about 300 comets recently detected
by the SOlar and Heliospheric Observatory (SOHO). In a few cases bright
comets close to the sun can be observed with the naked eye (a total of
seven comets since 1880, e.g. the great
September comet C/1882 R1, comet Pereyra 1963 R1, and comet
Ikeya-Seki 1965 S1). Some of these comets are observed to split into 2-4
fragments during perihelion passage. Much fainter members of the Kreutz
comet group are detected by space-based coronographs. The Solar Maximum
Mission (SMM) and the SOLWIND coronograph detected a total of 16 Kreutz
comets between 1979 and 1989. The Large Angle and Spectrometric COronograph 
(LASCO) on SOHO, which is sensitive to still fainter objects, detected about
300 sungrazers between January 1996 and June 2001.
Marsden (1989) showed from a detailed analysis of the orbits of the sungrazers 
known at the time that all of them are results of successive fragmentations of a
common progenitor, maybe a bright comet which appeared in the year 371 BC.

None of the sungrazers discovered from space was observed after
perihelion. The perihelia of most of them are between one and
two solar radii, so they did not fall into the sun.
The only available size estimate of a sungrazer observed by SOHO is
from the Ly\,$\alpha$ intensity of C/1996 Y1 measured by the
UltraViolet Coronograph Spectrometer on SOHO (Raymond et al. 1998). Its
radius is only $\approx 3.5\,$m which means that objects with
a diameter of a few meters are detectable.

\pagebreak
We assume that the sungrazers observed from spacecraft are destroyed
completely during their passage near the sun. One might argue that a
comet which is depleted of its volatiles or covered by a refractory
crust during its perihelion passage might be difficult to detect
post-perihelion even if it survives its perihelion passage. But the
detection of Fe\,I and Ni\,I emissions in spectra of the great September
comet C/1882 R1 (Copeland and Lohse 1882) and the appearance of Na\,I, K\,I,
Ca\,I, Ca\,II, Cr\,I, Co\,I, Mn\,I, Fe\,I, Ni\,I, Cu\,I, and V\,I in
comet Ikeya-Seki at heliocentric distances below $0.2\,$AU (Preston 1967,
Slaughter 1969) suggests that even the dust component of the nucleus will
evaporate under the extreme conditions encountered by a sungrazing comet.
Therefore we conclude that the sungrazers discovered from space did indeed
not survive their perihelion passage.

This work is focused mainly on the question of what can be inferred
from the complete destruction. Two processes can contribute to the destruction
of the sungrazing comets detected by SOHO: sublimation and disruption. The
size of the surface layer which sublimates during the perihelion passage of
a sungrazing comet was first estimated analytically by Huebner (1967), who
equated the energy sublimated by the sungrazer with the solar input energy.
A more detailed numerical model which included thermal emission of the
nucleus and heat conduction into the nucleus was published by Weissman (1983).
Both models result in a few tens of meters for the size of the sublimating
layer. Disruption mechanisms have so far not been discussed in the context
of the destruction of sungrazing comets.

We calculate the maximum size of a sungrazer which is destroyed by
sublimation alone with a numerical model similar to that by Weissman. A nucleus
of pure water ice is assumed. We derive a semi-analytical approximation 
from our model which fits the numerical results from Weissman (1983) and
the present work well. The implications of the porous nature of cometary 
water ice on the thickness of the sublimating layer are discussed.

We then study if sungrazers larger than the size limits for destruction
by sublimation alone can be completely destroyed by a combination of
sublimation and disruption. As a first candidate, we study tidal disruption.
We show that the combination of sublimation and tidal disruption
allows a complete destruction of nuclei larger than
this limit only if sublimation is highly anisotropic. We also show that 
consideration of disruption by thermal stress does not change our upper limits 
significantly. Based on these results, implications for the Kreutz family
of sungrazers and the nucleus of the progenitor are discussed.

Finally we consider the possibility of an {\itshape in situ} detection at
$1\,$AU of ions from a sublimating sungrazing comet. We conclude that while
expected fluxes are sufficiently high to be measurable, the probability of a
spacecraft actually passing through such an ion cloud is only a few percent.

\pagebreak

% ******************************************************************
\section{Model description}
The model is restricted to the nucleus and does not include the
coma. We assume that the nucleus is spherical and rotating sufficiently fast
that all physical quantities are radially symmetric.
The chemical composition is assumed to be $100\,\%$
H$_2$O ice. At the surface, the energy balance
is expressed with
\begin{equation}
\label{eq:balance}
\frac{P_{\text{sun}}(1-A)}{16\pi d^2}=\sigma T_{\text{s}}^4
+Z(T_{\text{s}})\cdot L(T_{\text{s}})-F_{\text{s}}
\end{equation}
(e.g. Fern\'{a}ndez and Jockers 1983).
Here $P_{\text{sun}}\approx 3.83\cdot 10^{26}\,$W is the power of the sun,
$A$ the cometary albedo,
$d$ the heliocentric distance,
$\sigma$ the Stefan-Boltzmann-constant,
$T_{\text{s}}$ the surface temperature, 
$Z(T)=p(T)\cdot\sqrt{\mu/(2\pi T\, k)}$
the sublimation flux in kg$\,$m$^{-2}\,$s$^{-1}$
(as derived by Delsemme and Miller 1971; $p(T)$ is the vapor
pressure in thermodynamical equilibrium fitted by
Busch (1960), $\mu$ the mass of
one molecule and $k$ the Boltzmann \mbox{constant}),
$L(T)=-582\cdot T+2.62\cdot 10^6$ the latent heat of sublimation
in J$\,$kg$^{-1}$ which was obtained by fitted data
(Auer 1961) and
$-F_{\text{s}}$ the net heat conduction flux towards the
interior of the nucleus. With an initial temperature
distribution in the nucleus (the nucleus is assumed
to be at its radiative equilibrium temperature far from the sun) and the initial
heliocentric  distance, this equation provides $F_{\text{s}}$. 

The surface temperature depends on the heat conduction process in
the interior of the nucleus. In order to simulate this process,
the nucleus is divided 
into $l$ radial layers of equal volume (see Fig. 1; temperatures are 
cell-centered while heat fluxes are defined at cell edges). The heat flux $F_i$
between the layers $i$ and $i+1$ is calculated from
$$
F_i=-K(T(R_i))\cdot\frac{T_{i+1/2}-T_{i-1/2}}{\Delta r_i}.
$$
Here
$T(R_i)$ is the temperature interpolated at the position $R_i$
corresponding to $F_i$,  $\Delta r_i$ the distance between
the positions of $T_{i-1/2}$ and $T_{i+1/2}$, and $K(T)$ the heat conductivity 
of the ice in W$\,$K$^{-1}\,$m$^{-1}$. The heat conductivity is not very well
known for cometary ice and is one of the variables in the model. The temperature
at the layer boundaries and the $\Delta r_i$ are calculated by linear 
interpolation:
$$T(R_i)=T_{i-1/2}+\frac{T_{i+1/2}-T_{i-1/2}}{R_{i+1}-R_{i-1}}\cdot\left(R_i-R_{i-1}\right)$$
and
$$\Delta r_i=\frac{1}{2}\cdot\left(R_{i+1}-R_{i-1}\right).$$
The upper boundary flux $F_l$
is equal to $F_{\text{s}}$, the lower $F_0=0$ because of
radial symmetry. The heat fluxes allow to simulate the
heat diffusion process by calculating the
time derivative of the temperature in each layer, 
obtained from energy conservation:
$$
\dot{T}_{i-1/2}=\frac{-4\pi R_i^2\cdot F_i+4\pi R_{i-1}^2\cdot F_{i-1}}
{V_i\cdot\varrho\cdot C_p(T_{i-1/2})}
$$
with $V_i$ the volume of layer $i$ and
$C_p(T)=16.1\cdot T^{0.871}-39.2$ the
heat capacity of ice in J$\,$kg$^{-1}\,$K$^{-1}$ obtained from a fit
to data (Hobbs 1974, Auer 1961).
\pagebreak

Connecting this with an orbit integrating routine,
the model provides among other quantities the nucleus size $R$, the
sublimation rate $\dot{m}\equiv -Z\cdot 4\pi R^2$, and the radial
temperature profile of the nucleus as a function of time or true
anomaly.

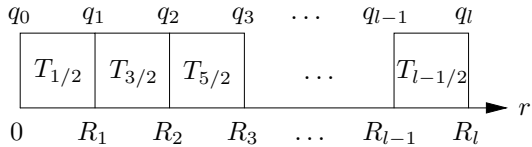
\begin{figure}
\label{fig:layers}
   \begin{center}
     \input{fig1.eepic}\\[7mm]
     \caption{Layer geometry}
   \end{center}
\end{figure}

% ******************************************************************
\section{Model test: $\dot{m}$ of comet P/Halley}
As a model test we calculate the mass of water ice sublimating 
from the surface of comet P/Halley. The only free parameter in the model test
is the thermal conductivity of the cometary material. 

The nucleus of comet P/Halley has approximately the shape of a prolate ellipsoid
with semi-axes [km] $8\times 4\times 4$ (Whipple 1987). A sphere with the same 
surface has a radius $R_0$ of $\approx 5000\,$m. Measurements during the Giotto
mission showed that $A\approx 0.04$. We assume a density of 
$\varrho = $931 kg m$^{-3}$  (the density of crystalline ice at 150 K). Some of
our sungrazer models will use a lower density of 600 kg m$^{-3}$. However, we 
perform our model tests for a single density only because the sublimating mass 
$\dot{m}\equiv -Z(T_S)\cdot 4\pi R^2$ does not depend strongly on density. 
Although the change in surface temperature with time and therefore $Z$ varies 
somewhat with  $\varrho$, the effect is much weaker than the dependence of 
temperature on heat conductivity. Since the total mass sublimated during one
orbital period is given by $m \approx 4\pi R_0^2\, \Delta R\, \varrho$, 
independence of the sublimated mass on $\varrho$ means that $\Delta R$, the 
radius of the layer which is sublimated during one orbit, is 
approximately proportional to $1/\varrho$. 

In  Fig. 2 we show the results of our model using
the parameters discussed above. The orbit of comet P/Halley is shown in the 
form of curves for $\dot{m}$ as functions of the true anomaly $\vartheta$. We 
varied the heat conductivity between 0 and the values for crystalline ice (taken
from Hobbs (1974) and Bosnjakovic (1972) and shown in Fig. 3  
as a function of temperature). The model results are compared to actual 
measurements. The theoretical curves fit through the measurements, which 
themselves scatter by up to half an order of magnitude. The three empty circles
at $\vartheta=0$ are peaks of the model curves (with one free parameter
$n=0,1,2$) as derived by Peale (1989) from the lightcurve of P/Halley scaled 
with the sublimation rate at spacecraft encounter.

\begin{figure}
 \label{fig:sublimrate}
 \begin{center}
     \epsfig{file=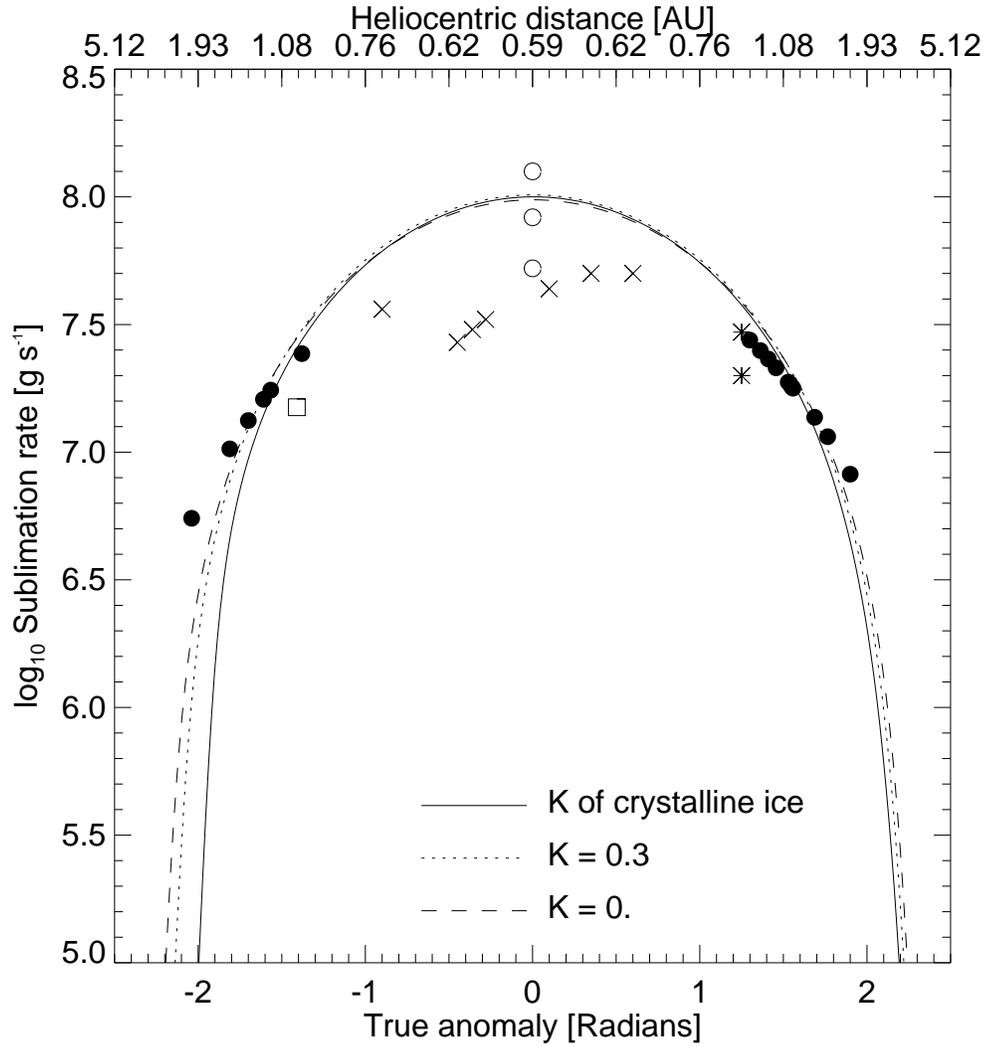}
 \end{center}    
\caption{Comparison of the sublimation rate of comet P/Halley between our model
calculations for various heat conductivities and several measurements. The model
radius of the comet is 5000\,m. The temperature dependent heat conductivity of
crystalline ice was taken from Bosnjakovic (1972) and Hobbs (1974). 
Measurements:}
$\star$: in situ estimates at Giotto fly-by from
Krankowski and Moroz (Peale 1989);
$\times$: estimates from Ly $\alpha$ intensities measured by the
Pioneer Venus spacecraft, Stewart 1987 (Peale 1989);
empty circles at $\vartheta=0$: peaks of the
model curves ($n=0, 1, 2$) calibrated by the sublimation rate at
spacecraft encounter (Peale 1989);
solid circles: Gas mass release rate (Singh 1992);
$\Box$: Swamy 1990
\end{figure}

\begin{figure}[tb]
\label{fig:heatcon}
\begin{center}
 \epsfig{file=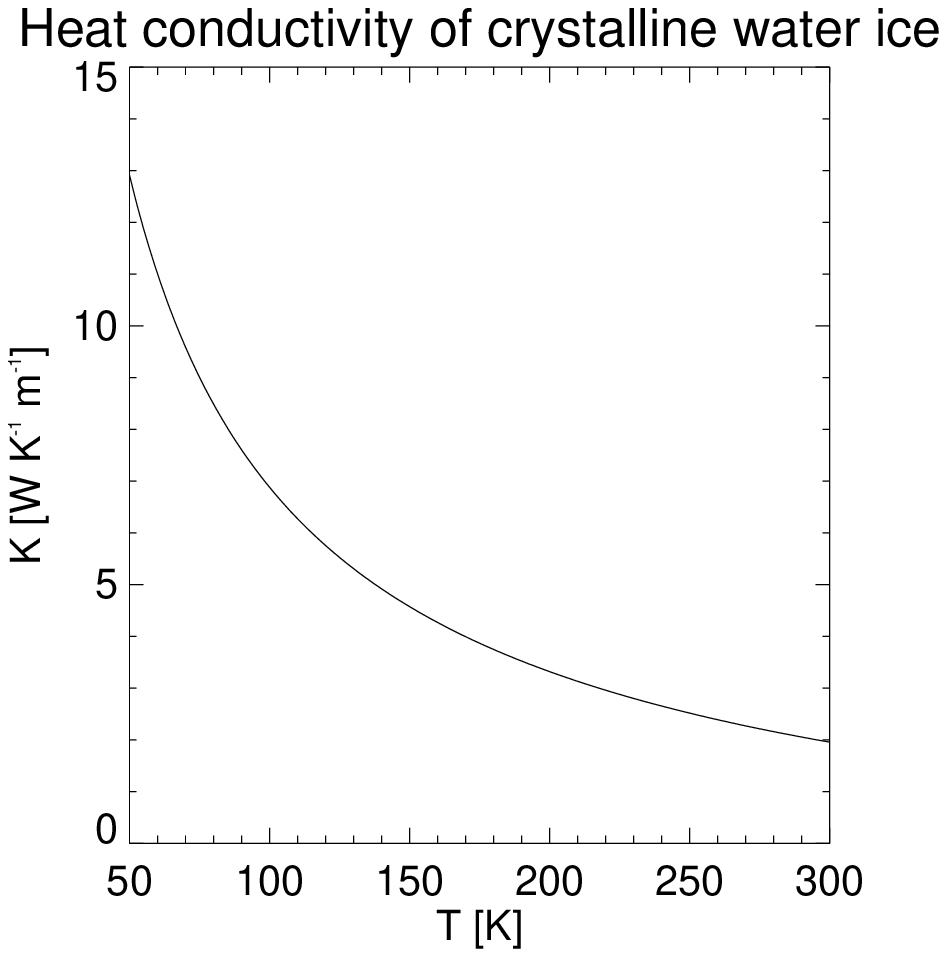}
\caption{ Heat conductivity of crystalline ice as a function of
          temperature. After Hobbs (1974) and Bosnjakovic (1972).}
 \end{center}	  
\end{figure}

Some discrepancies between model and observations exist. The model
predicts the production rate at perihelion to be higher than the
observed values. This can be readily explained by P/Halley not 
consisting of water ice only and by the observation that only
$\approx$\,10\,\% of Halley's surface is active (Keller et al. 1987).
Also the increase of the production rate is too steep in the model,
i.e. the onset of sublimation of water ice is too close to the sun.
This discrepancy decreases with decreasing thermal conductivity indicating
that the thermal conductivity of cometary water ice is much smaller than that
of crystalline ice. A similar conclusion was drawn by Weissman (1987) and Julian
et al (2000). On the 
other hand, the dependence of the total sublimation on the thermal conductivity
is very weak. $\Delta R$ only varies between 1.79\,m and 1.92\,m in the 3 cases
shown in fig. 2.

Model tests for comet C/Hale-Bopp 1995 O1 show additional evidence
for a low thermal conductivity. The lower the thermal conductivity the earlier
is the onset of significant sublimation. This result is in agreement with
K\"uhrt (1999) who was able to explain the early onset of sublimation of 
Hale-Bopp  with a combination of a low thermal conductivity and the low 
inclination of the cometary spin axis relative to its orbital plane.

Although the model does not reproduce all aspects of the observations
and the comparison is hampered by the large scatter of the measurements,
we conclude that the simple model is a good first approximation, despite 
its limitations of assuming a spherical nucleus of pure H$_2$O ice.
\pagebreak

% ******************************************************************
\section{Destruction by sublimation alone}
\label{sec:first}
In this section, we calculate the maximum size of a comet destroyed by
sublimation alone as a function of the parameters $\varrho$, $q$ and $A$
which describe the comet in our one dimensional model. Other disruption
mechanisms will be discussed in later sections.

\subsection{Thickness of the sublimating surface layer of a sungrazer}
We now use our model to calculate the thickness of the layer which will 
sublimate during the perihelion passage of a sungrazer. We will consider
crystalline and porous water ice.

The thickness of the sublimated layer does not depend significantly on the 
radius of the comet. This can be seen from
\begin{equation}
\frac{dm}{dt} = 4 \pi R^2 \varrho \, \frac{dR}{dt} = Z \, 4 \pi R^2
\end{equation}
or
\begin{equation}
\frac{dR}{dt} = \frac{Z}{\varrho},
\end{equation}
where $\frac{dm}{dt}$ is the sublimation rate, $Z$ the sublimation flux, 
and $R$ the radius of the cometary nucleus. The sublimation flux and therefore
the change in radius do not depend explicitly on the size of the nucleus. There
is a small implicit dependence: A very small nucleus may be heated up
internally very rapidly and then heat conduction cannot transport energy inward
from the surface as efficiently as for a larger nucleus. Since our model as
well as previous sublimation models (e.g. Weissman 1983, K\"uhrt 1999) show that
comets are heated up to a depth of at most a few meters, this effect can be 
neglected for all practical purposes.

Fig.\,4  shows the thickness of the sublimated layer for a 
sungrazer consisting of compact water ice ($\varrho = $931 kg/m$^3$ and heat 
conductivity of crystalline ice as shown in fig. 3) as a function of 
heliocentric distance. The sublimation during a complete orbit is considered.
The upper size limit derived this way is too high. The disappearance of the 
comets implies that they do not reach a heliocentric distance of more than a few
solar radii post perihelion. However, the difference is not very large. For a 
typical sungrazer, only 20--25\,\% of the radius of the sublimated layer 
evaporates post perihelion at a heliocentric distance of more than 3 solar 
radii. Since the maximum heliocentric distance a sungrazer can reach after 
perihelion without being discovered is hard to define,
we conservatively define our upper size limits as the layer which evaporates 
during one full orbit. 

\begin{figure}[tb]
\label{fig:any_rotator}
\begin{center}
\epsfig{file=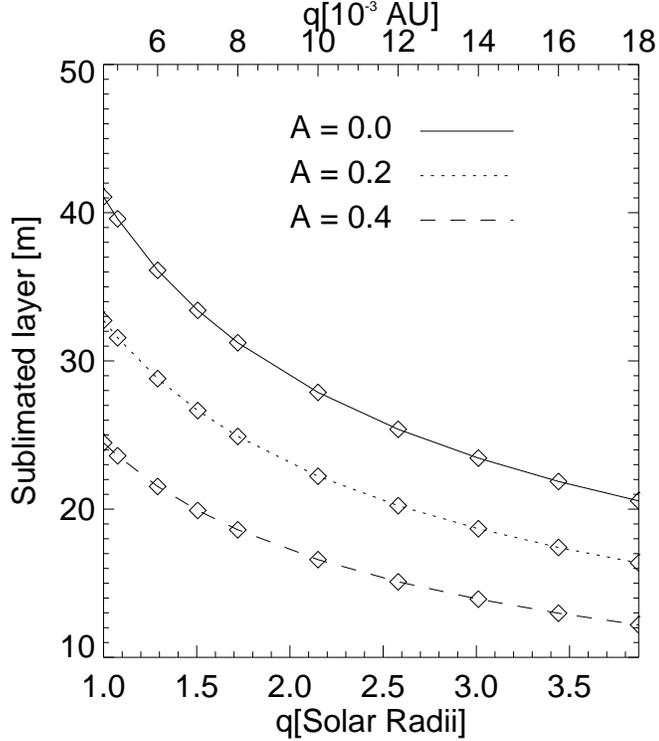}
\caption{Thickness $\Delta R$ of the layer sublimated from
a sungrazing nucleus of compact H$_2$O ice during the
complete perihelion passage.} 
\end{center}
\end{figure}

A comparison of fig. 4 with the results of the similar model of Weissman (1983) 
shows a maximum difference of $\approx 7\,\%$. While this shows the excellent
agreement between the two models, it is not an estimate of the error of the
models, because both use the assumption of a spherical nucleus composed of 
pure water ice.

The cometary H$_2$O ice may be porous and not crystalline which affects 
two quantities in the model: heat conductivity $K$ and density $\varrho$.
Heat conductivity is expected to be much lower in porous ice than in 
crystalline ice. Fig. 5  shows results of our model for 
various values of $K$. The variation of the thickness of the sublimated layer 
with $K$ is less than 3\,\%. Therefore the choice of $K$ is uncritical as long
as one is interested in the total sublimated mass only. In the model we use
$K$ = 0.15 W/(K m).

\begin{figure}[tb]
\begin{center}
\epsfig{file=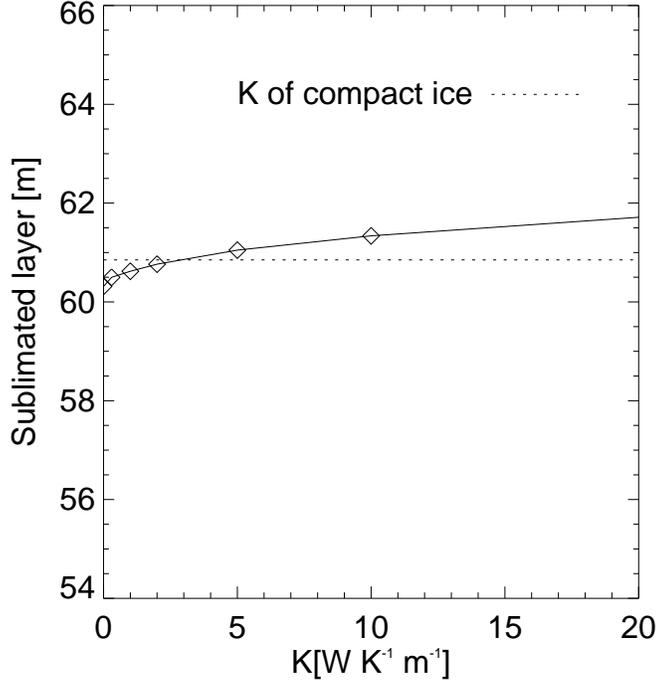}
\end{center}
\caption{Layer sublimated from a sungrazer during one perihelion passage as a
        function of the heat conductivity. The density is 600\,kg\,m$^{-3}$,
        the initial radius of the comet 200\,m, and the heliocentric distance
        at perihelion 0.005\,AU (1.1\,R$_{\text{sun}}$).}
\end{figure}

The density of porous ice is also lower than that of crystalline ice. Here
we use  the density of 600 kg/m$^3$ which Asphaug and Benz (1996)
derived for comet D/Shoemaker-Levy 9.  

Fig. 6  shows the thickness of the sublimated layer for a 
comet which consists of porous ice. The values are about 50\,\% higher than for
the comet made of crystalline ice and the maximum radius is about 60\,m.      

\begin{figure}[tb]
\begin{center}
\epsfig{file=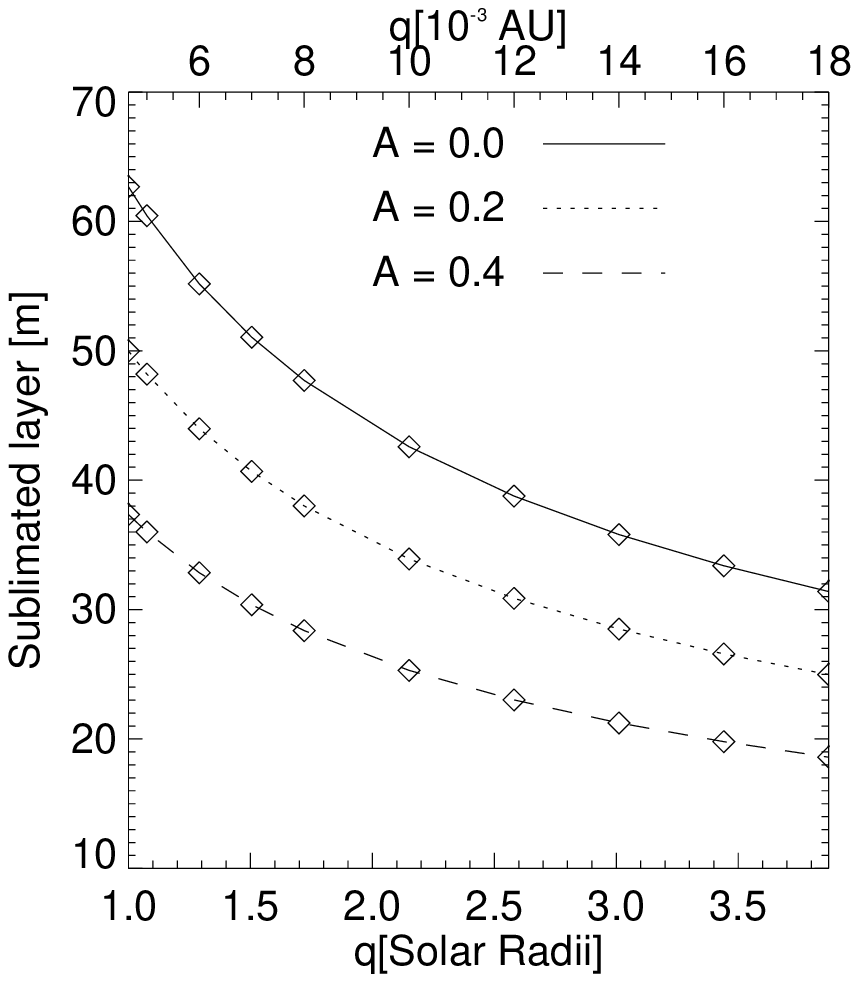}
\caption{Thickness $\Delta R$ of the layer sublimated from a sungrazing nucleus
         of porous H$_2$O ice (density 600\,kg\,m$^{-3}$, heat conductivity
         0.15 K$\,$W$^{-1}\,$m$^{-1}$) during the
         complete perihelion passage}
\end{center}
\end{figure}

\subsection{Semi-analytical approximation}
Comparing the radiation and the sublimation terms in Eq.
\ref{eq:balance} (Fig. 7)
shows that whenever the sublimation process becomes important,
it is orders of magnitude higher than the radiation.
If one is interested in the destruction of sungrazers by
sublimation, therefore mainly in the quantity $\dot{m}$,
the radiation can be neglected.

\begin{figure}[tb]
\label{fig:any_rotator2}
\begin{center}
\epsfig{file=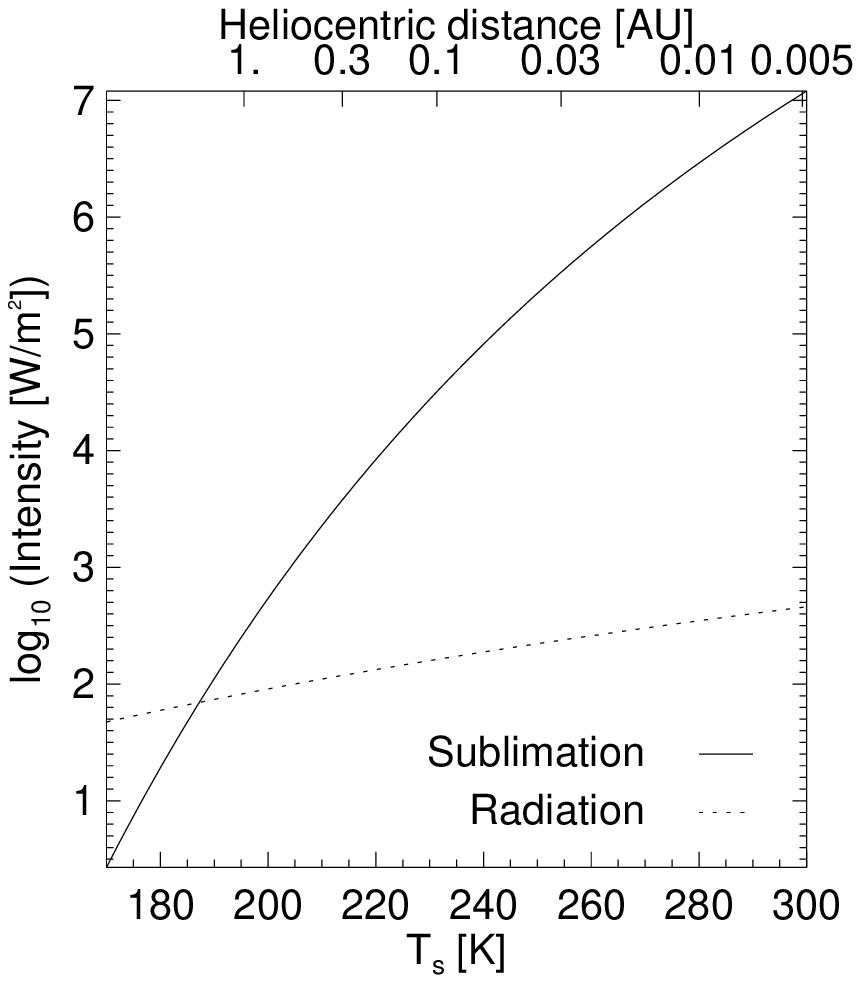}
\caption{Comparison of the radiation and the sublimation terms
        as functions of the surface temperature. The upper axis denotes the
        heliocentric distance of a sungrazer with perihelion at 0.005\,AU.}
\end{center}
\end{figure}

Model calculations with sungrazers
showed that at small heliocentric distances  $\approx 85\,\%$ of the
solar radiation energy goes into sublimation (Fig. 8).
Since only a negligible part of the sublimation takes place at large
heliocentric distances where sublimation is low and thermal radiation becomes 
important, we can simplify Eq. \ref{eq:balance} to
\begin{equation}
\label{eq:simple}
  \frac{P_{\text{sun}}(1-A)}{16\pi d^2}\cdot 0.85=Z\cdot L
\end{equation}
For H$_2$O, the latent heat is nearly independent of the
temperature, allowing to consider $L$ as a constant.
Replacing $Z$ by $-\dot{m}/(4\pi R^2)$ and solving for $\dot{m}$
yields
\begin{equation}
\label{eq:mdot}
  \dot{m}=-\frac{P_{\text{sun}}\cdot(1-A)}{4L\cdot d^2}
          \cdot R^2\cdot 0.85
\end{equation}
If $\dot{m}$ is substituted with $4\pi\varrho R^2\cdot\dot{R}$,
then solving for $\dot{R}$ yields:
\begin{equation}
\label{eq:Rdot}
  \dot{R}=-\frac{P_{\text{sun}}\cdot(1-A)}{16\pi\varrho L\cdot d^2}
          \cdot 0.85
\end{equation}
For parabolic sungrazer orbits, the time derivative of the true
anomaly is $\dot{\vartheta}=\sqrt{2GM_{\text{sun}}q}/d^2$
with $G$ the gravitational constant and
$M_{\text{sun}}\approx 1.989\cdot 10^{30}\,$kg the solar mass. Dividing
Eq. \ref{eq:Rdot} by $\dot{\vartheta}$ yields
\begin{equation}
\label{eq:dRdtheta}
  \frac{dR}{d\vartheta}=-\frac{P_{\text{sun}}\cdot 0.85}{16\pi\cdot\sqrt{2GM_{\text{sun}}}}\cdot\frac{1-A}{\varrho L\sqrt{q}}
\end{equation}
The rate of change in radius is independent of the position
on the orbit. This is a consequence of the sublimated energy being a constant
fraction of the solar input and is valid only when large heliocentric distances
are not important. Integration of eq. \ref{eq:dRdtheta} over the entire orbit 
yields the thickness 
$\Delta R:=|\int_{-\pi}^{+\pi}\frac{dR}{d\vartheta}d\vartheta|$
of the sublimated layer:
\begin{equation}
\label{eq:deltaR}
\boxed{
  \Delta R=\frac{P_{\text{sun}}\cdot 0.85}{8\cdot\sqrt{2GM_{\text{sun}}}}\cdot\frac{1-A}{\varrho L\sqrt{q}}
}
\end{equation}
It is important to note that $\Delta R$ is independent of $R_0$.
Collecting all comet-independent numerical factors and
fixing $L\approx 2.5\cdot 10^6\,$J$\,$kg$^{-1}$ yields
\begin{equation}
\label{eq:dRsimple}
\Delta R\approx 10^9\cdot\frac{1-A}{\varrho\sqrt{q}},
\end{equation}
where $\Delta R$ and $q$ are in meters and $\varrho$ in kg/m$^3$.
A similar result was obtained by Huebner (1967).

The comparison of the results of the numerical model with
Eq. \ref{eq:deltaR} shows a maximum discrepancy of 4\,\%.
Within the accuracy of the model, Eq. \ref{eq:deltaR}
can clearly be used to calculate $\Delta R$. The advantage of the analytical
approximation for future applications is that Eq. \ref{eq:deltaR} is much easier
to use than the full numerical model. 

\begin{figure}[tb]
   \begin{center}
   \input{fig8.tex}
 \caption{Ratio of the energy lead into sublimation to the total
        solar radiation energy}
 \end{center}
\end{figure}
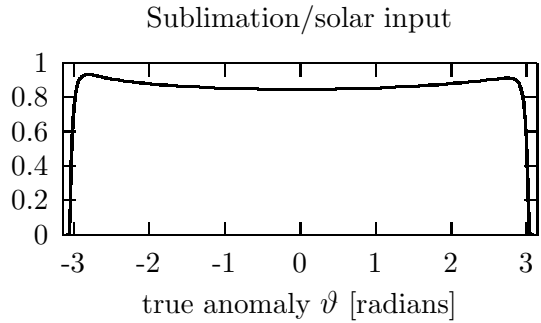
\pagebreak

\subsection{Discussion}
Given that we assumed that
a sungrazer is destroyed by sublimation alone,
its initial radius $R_0$ cannot exceed $\Delta R$. Hence
Eq. \ref{eq:deltaR} allows to calculate maximum
initial sizes as a function of $\varrho$. $q$ is known
for a given sungrazer, and $A$ can be set to zero
as generally the cometary albedo is very low, resulting
in an upper size limit valid for any albedo.

Considering only sungrazers but no sun-impactors implies
$q\geq R_{\text{sun}}$ (solar radius, $\approx 6.96\cdot 10^8\,$m);
Inserting $A$ = 0 and $q$ = $R_{\text{sun}}$ in eq. 9
yields an upper size limit for all sungrazers with a given density:  
$$
  R_0\leq 3.8\cdot 10^4\cdot\varrho^{-1},
$$
where $\varrho$ is in kg$\,$m$^{-3}$ and $R_0$ in m.
For the density of porous ice, $\approx 600\,$kg$\,$m$^{-3}$, which
was derived for comet Shoemaker-Levy 9 in Asphaug and Benz (1996),
the upper limit in size for the comets is $\approx 63\,$m.
\pagebreak

% ******************************************************************
\section{Additional destruction mechanisms}
If sungrazers are larger than the above size limit for complete
destruction by sublimation alone ($R_0>\Delta R$),
then an additional mechanism is needed to explain the disappearance
of all SOHO sungrazers. Since the lifetime against disruption of comets
in general is much larger than one perihelion passage, we restrict possible
disruption mechanisms to effects that are expected to increase with decreasing
heliocentric distance.

Since most of the sungrazers are likely to pass within the
Roche limit of the sun (Asphaug and Benz (1996) give
$R_{\text{Roche}}=1.51\cdot\sqrt[3]{M_{\text{sun}}/\varrho}
=3.23R_{\text{sun}}$ for $\varrho=600\,$kg$\,$m$^{-3}$),
tidal breakup is an obvious candidate mechanism. We show in the next
subsection that the combination of sublimation and tidal breakup
increases the upper size limits derived for destruction by sublimation alone if
and only if sublimation is highly anisotropic.

A second mechanism which may favor very low heliocentric distances is 
disruption by thermal stress. The second subsection shows that thermal
stress may separate small pieces from the comet but will be insufficient
to completely disrupt a large body.  
\subsection{The combination of sublimation and tidal breakup}
In this subsection we first discuss the isotropy of the sublimation from
the cometary nucleus. Since with the existing data it is difficult to evaluate 
to what extent sublimation from sungrazers is non-isotropic, we then treat
tidal disruption for the two extreme cases of isotropic sublimation and 
completely non-isotropic sublimation (no activity on the nightside of the 
comet). We show that for all realistic cometary densities
the combination of isotropic sublimation and tidal breakup is insufficient
to destroy a comet larger than $\Delta R$ during one perihelion
passage. We calculate upper size limits for the combination of sublimation and
tidal disruption in the case of completely non-isotropic sublimation. 

\subsubsection{Isotropy of sublimation from a cometary nucleus close to the sun}

Isotropy of sublimation is an important parameter in the treatment of tidal
breakup. In case of isotropic sublimation, the reaction force
of the sublimating molecules on the comet causes a compression which acts
against tidal stress and may prevent a comet from tidal disruption. While
sublimation is a surface effect, the resulting compression is felt throughout 
the body. The response time of the comet to an increase in surface pressure is
of the order $\tau$ = $R/c_s$, where $R$ is the radius of the cometary nucleus
and $c_s$ the speed of sound in water ice. For compact water ice, $c_s$ =
$\sqrt{E/\varrho}$, where $E$ is the Young modulus of the ice and $\varrho$
its density. For $E \approx$ 10$^{10}$ Pa (Hobbs 1974) and 
$\varrho$ =  1000\,kg m$^{-3}$, the sound speed is about 3\,km/s, resulting in
a response time of 0.3\,s for $R$ = 1\,km. For porous ice, the propagation of
pressure is more complicated, but sound speeds of the same order of 
magnitude as for compact water ice have been measured for materials which are
used as cometary analogs in simulation experiments (Kohl et al. 1990). Therefore
the propagation of the compression through the nucleus of the comet is fast
compared to changes in sublimation flux. In case of completely non-isotropic 
sublimation, there is no sublimation from the night-side of the nucleus and
therefore no compression. The reaction force from molecules sublimating on the 
dayside acts as the well known non-gravitational force which in this case 
pushes the comet away from the sun.

In our model sublimation is isotropic as a consequence of the fast rotator
approximation, i.e. the assumption that the cometary rotation is fast
compared to the time scale of changes in the solar input flux. Close to 
perihelion the modeled sublimation rate of a sungrazer changes by an order of 
magnitude within 2--3 hours. Typical rotation rates of comets are between 5 and
20 hours (Jewitt 1992). Although sungrazing comets may rotate faster than
other comets because they could have gained rotational energy in there history 
of multiple tidal disruptions from a common progenitor, it is possible that the 
fast rotator approximation is invalid close to perihelion.

If the cometary rotation is slow compared to changes in solar input, sublimation
will be highly non-isotropic unless the coma of the comet is optically thick.
Salo (1988) has used a Monte-Carlo
model of scattering in the coma of a comet to calculate the spatial 
distribution of the incoming sunlight on a comet as a function of the optical 
depth of the coma. We now try to estimate the optical depth of the coma of a
sungrazer close to perihelion in order to evaluate if significant sublimation
from the night side will take place.

The optical depth $\tau$ can be written as:
\begin{equation}\label{tau}
\tau  = \int^{s_{max}}_{s_{min}} N(s) \sigma_c(s) ds
\end{equation}
Here $N(s)$ is the column density of particles with a size between $s$ and
$s + ds$ and $\sigma_c(s)$ their scattering cross section. $s_{\text{min}}$ and
$s_{\text{max}}$ are the minimum and maximum grain size, respectively. For
simplicity, we assume that the outflow of the dust particles is isotropic and 
with constant velocity, and that their size distribution is independent of the 
distance from the nucleus. Then the column density is
\begin{equation}\label{Ns}
N(s) = \int^\infty_R  f(s) \frac{R^2}{r^2} dr = R\, f(s)
\end{equation}
where $f(s)$ is the number density at the cometary surface of particles with a 
radius between $s$ and $s + ds$, and $R$ the radius of the nucleus. 
\pagebreak

The 
scattering cross section is approximated by Rayleigh scattering for small 
particles ($<600$\,nm) and taken as the geometric cross section of the particle
for larger particles:
\begin{equation}\label{fs}
 \sigma_c(s) = \left\{ \begin{array}{l@{\quad}l} \pi s^2 & (s > s_f = 600\,nm)  \\
 \pi\frac{s^6}{s_f^4} & (s < s_f = 600\,nm) \end{array} \right. 
\end{equation}        

We take the particle size distribution $f(s)$ from {\itshape in situ} 
measurements in the coma of comet P/Halley (Mazets et al. 1986):
\begin{equation}\label{Mazet}
 f(s) = \left\{ \begin{array}{l@{\quad}l} 
c_1 s^{-2}    & (s < s_0 = 600\,nm) \\  
 c_2 s^{-2.75} & (s_0 < s < s_1 = 6000\,nm) \\
c_3 s^{-3.4}& (s > s_1) \end{array} \right. 
\end{equation}   

Three conditions are needed to determine the three constants $c_1$, 
$c_2$, and $c_3$. Two of these conditions are that eq. \ref{Mazet} is 
continuous at $s_0$ and $s_1$. The third condition is that the mass of 
all dust particles on the surface as expressed by $f(s)$ must be equal to that
described  by the ratio between sublimation flux and velocity:
\begin{equation}\label{norm}
\int^{s_{\text{max}}}_{s_{\text{min}}} f(s) \frac{4}{3} \pi s^3 \varrho \; ds = 
\frac{Z_d}{v}
\end{equation}

Here $\varrho$ is the density of the dust particles, $Z_d$ the dust production
rate, and $v$ the outflow velocity of the dust particles. 

Now eq. \ref{tau} can
be solved easily using eqs. \ref{Ns}--\ref{norm}. We consider 2 cases: 
Comet P/Halley at perihelion and a sungrazing comet with a radius of 100\,m. In
both cases we have taken the parameters $\varrho = 1000 $\,kg\,m$^{-3}$ (we 
assume the dust to be somewhat denser than the comet in general), 
$s_{\text{min}} = 0$  (a choice of little importance because the contribution of 
small Rayleigh scatterers to the total optical depth is very small), and 
$s_{\text{max}} = 1$\,mm. It should be noted that $s_{\text{max}}$ is the maximum 
particle size of the {\itshape visible} dust because determination of the 
production rate of the dust $Z_d$ from the sublimation model requires 
knowledge of the dust/gas-ratio. This ratio is determined either from 
remote observations which are not sensitive to dust particles which are much 
larger than the wavelength of the observations or by {\itshape in situ}
measurements which register particles smaller than a mm only. There might be an 
additional component of unobserved large dust particles which cannot be included
in our estimate. We assumed that half of the production rate in our model is gas
production and the other half is dust which is carried away by the gas, 
corresponding to a typical cometary dust/gas ratio of 1. 

For comet P/Halley we use a radius of 5000\,m, an outflow velocity of
500 m/s, and a dust production rate of $1.5\times 10^{-4}$ 
kg\,m$^{-2}$\,s$^{-1}$, corresponding to half the total mass loss at 
perihelion as derived in section 3. The resulting optical depth is 
$\tau = $ 0.02, somewhat lower than estimates of the global optical depth on 
comet Halley (0.05, Thomas and Keller (1990)), but of the same order of 
magnitude.
\pagebreak

For the sungrazer, we assume that the radius is 100\,m, the outflow velocity
1000\,m/s, and a dust production rate of 0.5\,kg\,m$^{-2}$\,s$^{-1}$, which
again is half the sublimation rate as calculated for a typical sungrazer close
to perihelion. The resulting optical depth is 0.7. From fig. 3--5 in Salo (1988)
 we estimate the minimum night side flux at $\tau = 0.7$ to be about
10\,\% of the maximum day side flux.  

While the absolute values of $\tau$ are highly uncertain because of the
uncertainties in some of the parameters we used (e.g. $s_{\text{max}}$, $f(s)$) 
and because of our simplifying assumptions (especially isotropic dust production
and negligence of dust evaporation and fractionation), it seems remarkable
that the optical depth in the coma of a comparatively small sungrazing comet may
be more than an order of magnitude larger relative to  that in the coma of a 
large comet at moderate heliocentric distance. The reason is that the difference
in production rate at perihelion, which  is more than 3 orders of magnitude 
higher for the sungrazer than for Halley, more than outweighs the effect of the
difference in size.

Because of the large uncertainty in the absolute value of $\tau$  
it is difficult to establish firm limits on the non-isotropy of the 
sublimation of a sungrazing comet close to perihelion. Therefore we will
now consider the destruction of sungrazers by a combination of sublimation and
tidal disruption for the two extreme cases of isotropic sublimation and of
no sublimation pressure at the nightside of the comet.

\subsubsection{Tidal breakup in case of isotropic sublimation}
Most favorable for tidal breakup is the assumption of a comet
without tensile strength: a rubble pile. Such a model has indeed been
suggested for comets (e.g. Weissman 1986). In this case,
the only processes working against tidal disruption are
the self-gravity of the comet and the sublimation pressure.
For $\varrho=600\,$kg$\,$m$^{-3}$ and $R=200\,$m,
the strength due to self-gravity, $G\varrho^2 R^2$
(Sridhar and Tremaine 1991), is about $1\,$Pa. For comparison,
the isotropic sublimation pressure in our model is 
$|\dot{m}|\cdot v/(4\pi R^2)=Z\cdot v$ with $v$ the thermal velocity of the 
sublimated molecules. For temperatures around perihelion 
($T\approx 280\,$K), $Z$ is about $1\,$kg per m$^2$ and s, and 
$v\approx\sqrt{8kT/\pi\mu}$, is about $600\,$m$\,$s$^{-1}$.
This corresponds to a sublimation pressure of $600\,$Pa, which is
much larger than self-gravity.  

Neglecting self-gravity, the tidal stress $\sigma_T$ must exceed the 
sublimation pressure of 600\,Pa in order to be sufficient to disrupt the 
comet. The tidal stress is given by 
$\sigma_T(d)=GM_{\text{sun}}\varrho R^2/d^3$
(Asphaug and Benz 1996), so that it reaches its maximum value for 
$d=R_{\text{sun}}$. We define the critical radius of the nucleus $R_{crit}$
as the radius for which tidal stress is just sufficient to disrupt the comet:

\begin{equation}
\label{eq:disr1}
R_{crit} =\sqrt{\frac{600\,\cdot
R_{\text{sun}}^3}{GM_{\text{sun}}\varrho}}=3.9\cdot
10^4\cdot\varrho^{-\frac{1}{2}},
\end{equation}
where $\varrho$ is in kg$\,$m$^{-3}$ and $R_{crit}$ in m.
If the nucleus is smaller than $R_{crit}$, tidal stress is too weak to destroy
the comet against the compressive force of sublimation pressure.
$R_{crit}$ is very large compared to $\Delta R$, therefore the decrease of
the cometary radius by sublimation of the surface layer on the comet's way to
perihelion can be neglected here. 
\pagebreak

We now show that a comet with radius $>R_{crit}$ cannot break up in boulders
which are sufficiently small to be sublimated completely. In order to destroy 
the comet completely by sublimation and tidal breakup, the largest remnant 
after the disruption must be smaller than $\Delta R$. Hence the ratio $f$ of
the mass of the largest remnant to the
mass of the nucleus before the breakup is limited by
\begin{equation}
\label{eq:disr2}
f<\left(\frac{\Delta R}{R_{crit}}\right)^3
\end{equation}
$\Delta R$ is given by Eq. \ref{eq:dRsimple} and limited by
\begin{equation}
\label{eq:disr3}
\Delta R\leq 10^9\cdot\frac{1}{\varrho\sqrt{R_{\text{sun}}}}
=3.8\cdot 10^4\cdot\varrho^{-1}
\end{equation}
Substituting $R_{crit}$ and $\Delta R$ to Eq. \ref{eq:disr2} from
Eqs. \ref{eq:disr1} and \ref{eq:disr3} yields
\begin{equation}
\label{eq:disr4}
f\leq 0.915\cdot\varrho^{-\frac{3}{2}}
\end{equation}
In Asphaug and Benz 1996, the ``normalized periapse''
$b:=q/R_{\text{Roche}}$ necessary for a tidal disruption
of a rubble pile with largest remnant mass fraction $f$,
is given as a function of $f$:
\begin{equation}
\label{bf}
b(f)\approx
\frac{1}{9}+\frac{3}{5}\cdot\sqrt{f}+\frac{1}{6}\cdot\left(f^{12}-f^3\right)
\end{equation}
From the monotonous increase of $b(f)$ follows that
the upper limit Eq. \ref{eq:disr4} for $f$ implies an
upper limit for $b$:
$$
b\leq b(0.915\cdot\varrho^{-\frac{3}{2}})
$$
This yields an upper limit for $q=b\cdot R_{\text{Roche}}$.
On the other hand, we have the restriction $q\geq R_{\text{sun}}$:
\begin{equation}
\label{eq:disr6}
R_{\text{sun}}\leq q=
b\cdot R_{\text{Roche}}\leq
b(0.915\cdot\varrho^{-\frac{3}{2}})\cdot 1.51\cdot\sqrt[3]{M_{\text{sun}}/\varrho}
\end{equation}
Solving Eq. \ref{eq:disr6} numerically for $\varrho$ results in
$$
\varrho\leq 55\,\frac{\text{kg}}{\text{m}^3}
$$
This value is probably much lower than the density of cometary nuclei, although
there are very few determinations of densities of comets. After the fly-bys of
five spacecraft at comet P/Halley several attempts were made to determine 
the density of the comet (Rickman 1986, Rickman 1989, Sagdeev et al. 1988). The
results were contradictory, probably because uncertainties in the mass 
determination were so high that no significant constraints on the density of 
P/Halley could be made (Peale 1989). The only reliable density estimates of a 
comet were derived from the breakup of comet Shoemaker-Levy 9 and result in 
approximately 600 kg m$^{-3}$ (e.g. Asphaug and Benz 1996). Since this is more 
than an order of magnitude above our upper limit, we conclude that in the
case of isotropic sublimation a sungrazer larger than the size limit for 
destruction by sublimation alone cannot be destroyed completely by a combination
of sublimation and tidal breakup.

\subsubsection{Tidal breakup in case of completely non-isotropic sublimation}

We first note that non-isotropic sublimation has no major consequences on the 
thickness of the sublimated layer derived in section 4: We showed 
that the energy balance close to the sun is dominated by solar input and 
sublimation. Since the solar input is independent on the rotation rate of the 
comet, conservation of energy requires that the sublimation rate will not be 
seriously affected by non-isotropic sublimation. 

For the treatment of tidal breakup the difference between isotropic and
non-isotropic sublimation is the absence of sublimation pressure in the 
latter case. Only self gravity and the tensile strength of the comet (if
present) work against disruption by tidal stress. This case is in many respects
similar to the situation which led to the disruption of comet Shoemaker-Levy 9 
during its close approach to Jupiter. The perihelion distance of the average 
sungrazer in solar radii (1.31 $R_{\text{sun}}$) is very close to the perijove 
distance of Shoemaker-Levy 9 to Jupiter in Jovian radii. Therefore some of the 
results about the splitting of comet Shoemaker-Levy 9 are valid for the analysis
of the disruption of sungrazing comets in absence of sublimation pressure at 
the night side.

In order to derive upper size limits for a comet which can be completely
destroyed by a combination of tidal disruption and sublimation, we again assume
a body without tensile strength. We use the results from 
Asphaug and Benz (1996) (reproduced as eqs. \ref{bf} and \ref{eq:disr2} in the 
previous subsection) to determine the maximum size of a strengthless comet which
may disrupt in a way that all fragments are smaller than the largest fragment 
size which will sublimate completely. Fig. 9 shows the 
results. If we assume (somewhat arbitrarily) that the minimum conceivable 
density of a comet is 200 kg m$^{-3}$, the maximum size before disruption would
be about 360\,m. 

\begin{figure}[tb]
\begin{center}
\epsfig{file=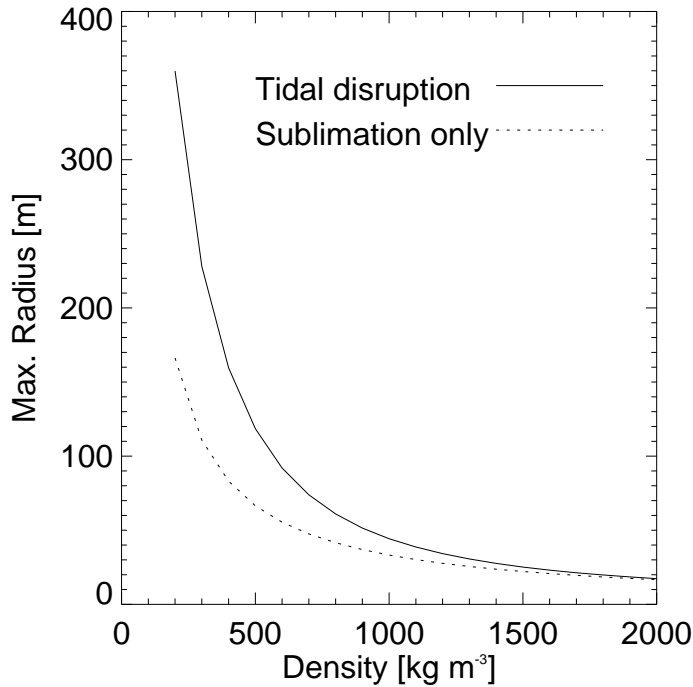}
\caption{ Maximum radius of a comet which will be destroyed by sublimation alone
        (dashed line) and by a combination of sublimation and tidal disruption
        (solid line). No tensile strength of the comet and no sublimation on
        the nightside is assumed.} 
\end{center}
\end{figure}

Finally, in case of non-isotropic sublimation it is possible that variations in 
sublimation pressure may induce a stress on the comet. The difference between
the pressure at the day side and at the night side is of low importance here as 
it acts on the comet as a whole. On the other hand, local differences in 
sublimation pressure may act to separate fragments from a non-spherical nucleus
(the reaction force may act on material close to the visual limb as seen from 
the sun). For most shapes of the nucleus such local pressure gradients result
in the separation of small fragments only. Therefore they are unlikely to
increase our upper size limits significantly. A more quantitative treatment of 
local pressure effects and their influence on the size of the sublimating layer
would require a full 3-dimensional model of the cometary nucleus which is beyond
the scope of the present paper.

\subsubsection{Summary}
The rapid perihelion passage of sungrazing comets may cause 
non-isotropic sublimation, although the unknown rotation rates of the sungrazers
and the uncertain optical depth of their coma make it difficult to estimate
the difference between the sublimation fluxes at the dayside and at the
nightside. Should sublimation be isotropic, a combination of tidal 
disruption and sublimation will not destroy a comet which is larger than
the size limits derived from destruction by sublimation alone. In case of
completely non-isotropic sublimation, the size and density limits are modified 
as shown in Fig. 9.

%
%\subsection{Other destruction mechanisms}
%In this paper, sublimation and tidal disruption are the only
%destruction mechanisms considered quantitatively.
%More complex processes are beyond the scope of this model, such as
%crust formation which could lead to high sub-surface pressures
%and then to sudden ruptures. But thermal stress
%(Tambovtseva and Shestakova 1999), for instance, can exceed
%the strength of comet nuclei, especially for small sizes and
%close passages near the sun.
%

\subsection{Disruption by thermal stress}
% Thermal stresses MPa with inclusions 100 MPa (Kuerth, Kuerth +  Tauber)
% Tends to make small pieces. But medium-sized (100m) maybe disruption in
% larger pieces.

Thermal stresses in comets are generally much larger than tidal forces,
even for close encounters between a comet and the sun. Shestakova and
Tambovtseva (1997) note that even in Comet D/Shoemaker-Levy 9 during
Perijove in 1992 thermal stress close to the surface exceeded the
tidal forces on the comet by 4 or 5 orders of magnitude. Therefore thermal
breakup needs to be considered as another possible destruction mechanism
for sungrazing comets.

The thermal stress on the surface layer of a comet introduced by
non-linear radial temperature variations $T(r)$ can reach several MPa close
to the surface of a comet on the orbit of P/Halley (K\"uhrt 1984, Tauber
and K\"uhrt 1987). This value can increase to several hundred MPa if there
are inclusions of a material with a different thermal
expansion coefficient in a background material (e.g. CO or CO$_2$ inclusions
in H$_2$O ice) and may be still higher when the sublimation temperature of
the included material is exceeded.

Thermal stress within the H$_2$O ice cannot completely disrupt a
sungrazing comet. The main difference between the thermal profile of a
sungrazer at perihelion and that of a comet at 1 AU is the higher surface
temperature of the sungrazer (in our case of pure H$_2$O ice about 300 K
versus 200 K, see Fig. 10 ). Because of the high surface 
temperature nearly all of the energy input goes into sublimation of the ice. 
This leads to a very steep temperature gradient immediately below the surface.
Below surface depth of tens of centimeters, the temperature gradient is smaller
close to the sun than it is at 1 AU. Therefore the thermal stress in a sungrazer
does not increase during perihelion passage except for a very small layer below
the surface. Since the hot surface material sublimates rapidly, there is no 
significant effect of thermal stress within the H$_2$O ice on the destruction
of the comet.

\begin{figure}[tb]
   \begin{center}
   \epsfig{file=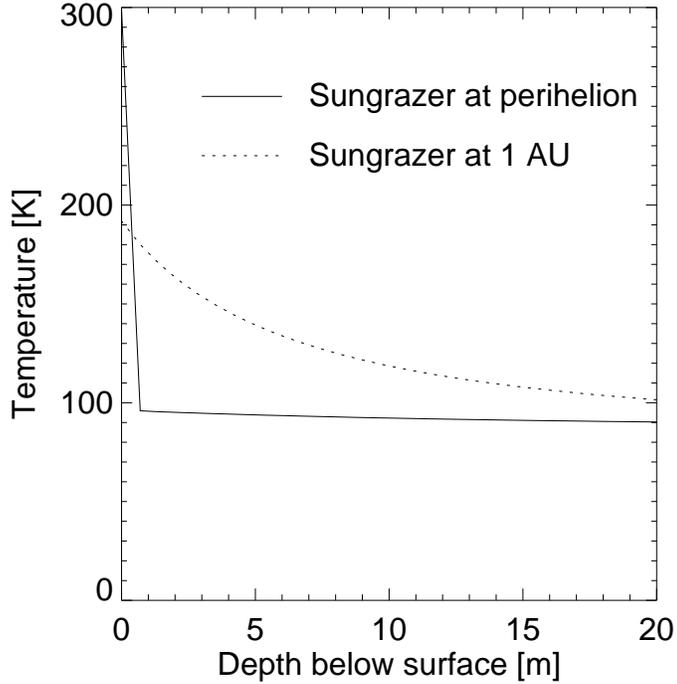}
\caption{ Temperature profile of a sungrazer at perihelion and at 1 AU. A 
        perihelion distance of 0.005\,AU (1.1\,R$_{\text{sun}}$ and the
        conductivity of crystalline ice was used.}
\end{center}
\end{figure}
Inclusions of e.g. CO or CO$_2$ in the H$_2$O ice may cause increased
thermal stress. Cracks in the ice may be induced and a block of cometary
material may be separated from the nucleus and accelerated by the gas
escaping from the inclusion. Tauber and K\"uhrt (1987) solve the rocket
equation to calculate the maximum
mass \textit{m$_f$} of a separated fragment which reaches escape velocity
from the cometary nucleus:
\begin{equation}
\label{inclusion}
m_f = m_i \frac{\exp{\left(\frac{v_e + g_0 t}{v}\right)}}{\exp{\left(\frac{v_e + g_0 t}{v}\right)}-1}
\end{equation}
Here \textit{m$_i$} is the mass of the inclusion, \textit{v$_e$} the escape
velocity from the cometary nucleus, \textit{v} the velocity of the escaping
gas, \textit{g$_0$} the gravitational acceleration at the surface of
the comet, and \textit{t} the time needed for complete sublimation of the
inclusion. Since \textit{v} increases and \textit{t} decreases as a comet
approaches the sun, the maximum fragment mass increases strongly during
the perihelion passage of a sungrazing comet.
\pagebreak
\renewcommand{\baselinestretch}{0.95}
\small\normalsize

From a comparison with minor fragments of observed cometary splittings,
Tauber and K\"uhrt (1987) estimate a typical radius of an inclusion of
\textit{r$_i$} = 0.5m. The sublimation time is determined from the mass
loss of the inclusion:
\begin{equation}
\frac{dm}{dt} = \frac{4}{3} \pi \varrho \, \frac{dr^3}{dt}
              = 4 \pi r^2 \varrho \, \frac{dr}{dt} = Z \, 4 \pi r^2
\end{equation}
Integration yields:
\begin{equation}
t = \frac{r_i \varrho}{Z}
\end{equation}

For a spherical inclusion with \textit{r$_i$} = 0.5m,
\textit{$\varrho$} = 600 kg/m$^3$, and
\textit{Z} = 3 kg/s (the model value for H$_2$O on a typical sungrazer at
perihelion; higher values are expected for inclusions of CO or CO$_2$), the
sublimation time is 100s. For the same density, the escape velocity
\textit{v$_e$} and gravitational acceleration \textit{g$_0$} are
5.8$\times$ 10$^{-4} \times $\textit{R}  m/s and
1.7$ \times$ 10$^{-7} \times $\textit{R}  m/s$^2$, respectively. Here
\textit{R} is the radius of the cometary nucleus. Since
\textit{g$_0$\,t} $\ll$ \textit{v$_e$} $\ll$ \textit{v}, eq. \ref{inclusion}
simplifies to:
\begin{equation}
m_f = m_i\frac{v}{v_e}.
\end{equation}

According to the sublimation model,
the thermal velocity of the molecule relative to the nucleus is
$\approx 0.5\,$km$\,$s$^{-1}$; this velocity is quite small
as the model does not include dust, which in reality leads to higher
surface temperatures and hence to higher vaporization
velocities than those derived with the model.
In order to get a maximum estimation for the radius
of the neutral cloud around the nucleus, the velocity corresponding
to the equilibrium temperature without sublimation,
$T_{\text{eq}}=\sqrt[4]{(1-A)\cdot P_{\text{sun}}/(16\pi d^2 \sigma)}$,
is used: $\approx 2\,$km$\,$s$^{-1}$.

For a nucleus with a radius \textit{R} of 63\,m (the upper size limit for
a comet which sublimates completely in case of isotropic sublimation), the mass
of the largest possible fragment is about 55000 times larger than that of the 
inclusion, but only 2.7 \% of the mass of the cometary nucleus. The mass 
fraction decreases to 0.4 \% for \textit{R} = 100\,m and to 4$\times$ 10$^{-7}$
for \textit{R} = 1\,km.

Therefore unrealistic large numbers of inclusions are necessary to remove
a large fraction of a comet larger than the size limit for destruction by
sublimation alone. Consideration of thermal splitting with or without
inclusions of different ices does not change significantly the upper size
limit for comets which disappear during perihelion passage.

\subsection{Other disruption mechanisms}
Other mechanisms like collisions, explosions of volatiles below the
surface, and fast rotation have been suggested as causes for cometary
disruptions (see overview in Shestakova and Tambovtseva 1997). Of these 
mechanisms, collisions  are too infrequent to destroy all $\approx$300 comets 
seen by spacecraft coronographs. Fast rotation of all fragments seems 
extremely unlikely, too.
Explosions may be possible, but given that sungrazers
at perihelion are heated up to a depth of less than a meter, it would be
hard to explain why explosions occur much more frequently during the
perihelion passage of a sungrazing comet than during the time comets spend 
further from the sun. The perihelion passage of a sungrazer lasts only a few 
hours or a few days. For a typical sungrazer orbit($q$ = 1.3 $R_{\text{sun}}$),
50\% of the sublimated material evaporates during 3 hours around perihelion.

We conclude that the size limits we derived for the destruction of
sungrazing comets by sublimation and possibly by tidal disruption are the 
actual upper size limits of these comets unless some unknown highly efficient 
destruction mechanism exists.

\renewcommand{\baselinestretch}{1.}
\small\normalsize
% ******************************************************************
\section{Oxygen ion fluxes from sublimating comets at 1 AU}
We now consider the possibility of detecting ions from disintegrating
comets. We calculate ion fluxes an observer would measure at 1 AU in
the ecliptic plane. Then we discuss the probability of a detection by
an ion detector on a spacecraft located at 1\,AU.

\subsection{General considerations}

For this work, all considerations are restricted to
the average sungrazer orbit, derived by averaging the preliminary
orbit elements of all sungrazers discovered till August 1999,
(Douglas Biesecker, personal communication).
\begin{table}[H]
   \begin{center}
   \begin{tabular}{ccrcrl}
$q$&$=$&$(1.31$&$\pm$&$0.26)$&\hspace{-3mm}$R_{\text{sun}}$\\
$i$&$=$&$(142$&$\pm$&$5)$&\hspace{-3mm}°\\
$\Omega$&$=$&$(-1$&$\pm$&$19)$&\hspace{-3mm}°\\
$\omega$&$=$&$(79$&$\pm$&$16)$&\hspace{-3mm}°
   \end{tabular}
   \end{center}
\end{table}
\noindent
where the symbols have their usual meaning.

This orbit is plotted in Fig. 11. An instrument in the ecliptic plane
will measure ions originating at the ascending note, which is at
2.19\, R$_{\text{sun}}$.

\begin{figure}[tb]
 \begin{center}
     \epsfig{file=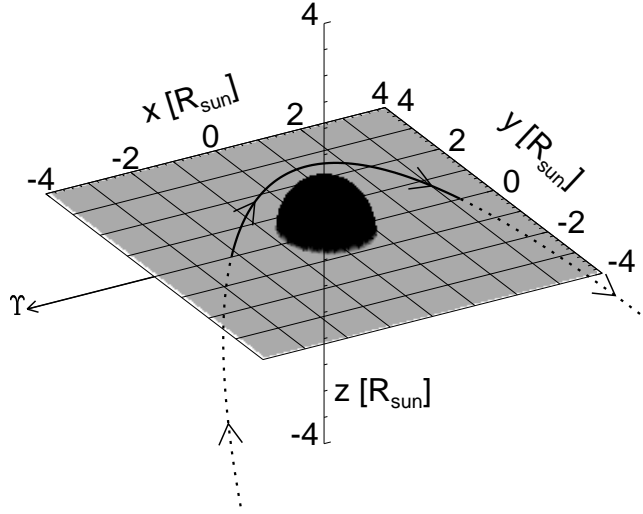}
 \caption{Averaged sungrazer orbit with $q=1.3R_{\text{sun}}$. The ecliptic 
plane, the sun and the direction of the first point of Aries are shown. 
The ascending node is close to the first point of Aries ($\Omega\approx 0$) 
on the negative $x$-axis. In spring, we see the sun in the first point of Aries,
so the earth is on the side of the descending node. Cometary ions can be
measured in autumn when the earth is on the side of the ascending node.}     
 \end{center}
\end{figure}

The H$_2$O sublimation model provides $\dot{m}$ along the orbit.
Since sungrazers pass the sun very closely, H$_2$O molecules
will be photo dissociated rapidly. Scaling the photodissociation
rate of $12.99\cdot 10^{-6}\,$s$^{-1}$ at $1\,$AU
(Budzien et al. 1994) with $d^{-2}$ shows an average lifetime of
$\approx 7\,$s at 2 solar radii. Values from the same publication
indicate that $\approx 85.3\,\%$ of all H$_2$O molecules are
destroyed by photodissociation:
$$
\text{H}_2\text{O}+\gamma\to\text{OH}+\text{H}
$$
As $\approx 95\,\%$ of the solar wind consists of hydrogen,
cometary hydrogen cannot produce a measurable ion flux
peak from the comet.
$63\,\%$ of the OH molecules are destroyed by photodissociation,
and this with an average lifetime of $\approx 33\,$s at 2 solar radii:
$$
\text{OH}+\gamma\to\text{O}+\text{H}
$$
The neutral oxygen atoms will be rapidly ionized by
photons or electron impact, and then follow the solar wind
radially away from the sun. Therefore, different oxygen ions
are the only particles originating from comet H$_2$O sublimation
which might be measured at 1AU. We restrict our analysis to ions arriving at
1 AU close to or in the ecliptic plane. The
slow speed solar wind has to be considered in this case.
In order to find the charge state distribution expressed
by the dimensionless occupation numbers $n_0(t), n_1(t), \dots, n_8(t)$
for each charge state, the following system of coupled
differential equations must be solved:
\begin{eqnarray*}
\dot{n}_0&=&n_e\cdot(-I_{0,1}\cdot n_0+R_{1,0}\cdot n_1)-r_\gamma\cdot n_0\\
\dot{n}_1&=&n_e\cdot(I_{0,1}\cdot n_0-[I_{1,2}+R_{1,0}]\cdot n_1+R_{2,1}\cdot n_2)+r_\gamma\cdot n_0\\
\dot{n}_i&=&n_e\cdot(I_{i-1,i}\cdot n_{i-1}-[I_{i,i+1}+R_{i,i-1}]\cdot n_i+R_{i+1,i}\cdot n_{i+1})\text{~~for }i=2\dots 7\\
\dot{n}_8&=&n_e\cdot(I_{7,8}\cdot n_7-R_{8,7}\cdot n_8)
\end{eqnarray*}
with the normalization condition $\sum_{i=0}^8 n_i=1$.
$n_e[$cm$^{-3}]$
is the electron density fitted as a function of $d$: 
$n_e(d)=2.34\cdot 10^{23}\cdot d^{-2}$ (Bochsler 2000).
$I_{i,i+1}[$cm$^3\,$s$^{-1}]$
is the electron impact ionization rate from state $i$ to state $i+1$.
It is calculated  with the method by Arnaud and Rothenflug (1985).
$R_{i,i-1}$ is the electron recombination rate from state $i$
to state $i-1$ (Nahar 1998).
$r_\gamma$ is the photoionization rate from neutral oxygen to
O$^+$ and assumed to vary with $d^{-2}$. Scaling with a rate
of $7.5\cdot 10^{-7}\,$s$^{-1}$ at $1\,$AU (Rucinski et al. 1996)
yields $r_\gamma=1.68\cdot 10^{16}\cdot d^{-2}$. Both $I_{i,i+1}$
and $R_{i,i-1}$ are functions of the electron
temperature $T_e$, which is fitted by a function of $d$:
$T_e(d\,[R_{\text{sun}}])=
5.5\cdot 10^{10}\cdot[5.49\cdot 10^{-5}
-(0.2+d)\cdot$e$^{-10d}]/d$ (Bochsler 2000).
With this, the system of differential equations can be evaluated
at a given perihelion distance $d$. Together with a velocity
profile for the slow solar wind, $v_{\text{sw}}(d)$ (Bochsler 2000),
the system can be integrated from a starting distance $d_0$ out to
1\,AU, yielding the probabilities $n_0,\dots,n_8$
for an O$^+$ ion to arrive at 1 AU in the corresponding charge
state.

Adding the lifetimes for the processes H$_2$O
$\to$ OH $\to$ O $\to$ O$^+$
shows that a H$_2$O molecule emitted from the nucleus at the ascending
node produces an O$^+$ ion after only $\approx 2$ minutes. Within
this time, the nucleus moves only $\approx 0.07$ solar radii.
Therefore, ions to be measured at a spacecraft in the ecliptic plane 
originate from a very small orbit part around the ascending node (the 
descending node is excluded as no sungrazer was observed after the passage).
Integrating the set of charge state equations beginning at the
ascending node ($d_0=2.19R_{\text{sun}}$) yields the distribution
in Table I.

\begin{table}[tb]
 \begin{center}
   \begin{tabular}{|c|c|c|c|}
      \hline
\rule[-0mm]{0mm}{4.5mm}    &   O$^{4+}$&O$^{5+}$&O$^{6+}$\\
\hline
Relative distribution &0.215&0.605&0.180\\
\rule[-0mm]{0mm}{4.5mm}Absolute ion flux & $1.0\cdot 10^{11}$&$2.9\cdot 10^{11}$
&$8.6\cdot 10^{10}$\\
\hline
   \end{tabular}\\[7mm]
\caption{Charge state distribution and ion fluxes in particles per m$^2$ and s
of the oxygen ions from a sungrazer with a nucleus of
$\varrho=600\,$kg$\,$m$^{-3}$, $A=0$ and initial radius $28\,$m.}
   \end{center} 
\end{table}

\subsection{Size of the cloud of sublimating material}
An instrument in the ecliptic plane could potentially measure ions
originating at the ascending node of the cometary orbit.
An H$_2$O molecule leaving the nucleus at the
ascending node of the average orbit has a lifetime of $8\,$s till
photodissociating to OH.
In order to get a maximum estimation for the radius
of the neutral cloud around the nucleus, we use the expansion velocity
corresponding to the equilibrium temperature without sublimation of
$\approx 2\,$km$\,$s$^{-1}$. The photodissociation
to OH leads to a velocity addition of $\approx 1\,$km$\,$s$^{-1}$
(Johnstone 1991, Wu and Chen 1993). On average, the velocity vectors
add up perpendicularly: $\sqrt{2^2+1^2}\,$km$\,$s$^{-1}$.

The lifetime for
OH is $\approx 40\,$s. The photodissociation to O again
leads to a velocity addition of $\approx 1\,$km$\,$s$^{-1}$
(Johnstone 1991). Finally, the lifetime for neutral oxygen is
given by $(r_\gamma+n_e\cdot I_{0,1})^{-1}$: $\approx 64\,$s.
Adding up the distances for the 3 phases shows that
the radius of the neutral cloud around the nucleus is
$\approx 2\,$km$\,$s$^{-1}\cdot 8\,$s$+\sqrt{5}\,$km$\,$s$^{-1}\cdot 40\,$s$+\sqrt{6}\,$km$\,$s$^{-1}\cdot 64\,$s $\approx 262\,$km
(maximum estimation for the average).

Once an oxygen atom has been ionized it will quickly lose its velocity relative
to the solar wind flow. The initial velocity perpendicular to the solar
magnetic field (which we assume to point radially away from the sun) is 
the cometary velocity component at 2.19\,$R_{sun}$ of $v_{perp} = 322\,$km/s. 
This will cause the ion to gyrate around a solar wind magnetic field line. For a
magnetic field $B$ of $9.6 \cdot 10^{-5}\,$T (corresponding to 10$\,$nT at 1 AU 
and an inverse square law dependence for $B$ as a function of heliocentric 
distance), the radius of gyration for O$^+$ ions at the ascending node is 
$r_g = (m_O v_{perp})/(e B) \approx 560\,$m, where $m_O$ and $e$ are the mass
and charge of the oxygen ion, respectively. The time scale for this
guiding-center pickup is about one gyration period $2\pi r_g/v_{perp}\approx
0.01$\,s, so that onset of gyration can be considered as instantaneous. Since
the gyration radius  is small compared to the size of the oxygen cloud, any 
motion perpendicular to the direction of solar wind expansion can be neglected.
Along the magnetic field lines newborn ions move with an initial velocity of
$260\,$km/s towards the sun. They are decelerated and brought to solar wind 
velocity ($\approx 100\,$ km/s directed away from the sun at 
$2.2 R_{\text{sun}}$) by outward propagating fluctuations in the solar magnetic
field (e.g. Gaffey et al. 1988). While a quantitative treatment of this process is beyond the scope
of the present paper, we expect acceleration to occur fast because of the high 
power of ion-cyclotron waves close to the sun. Hence in what follows we assume 
that the particles are captured by the solar wind immediately after ionization 
and move radially away from the sun with solar wind velocity instantaneously. 

The mass in the particle cloud, with a 
fraction of $16/18$ of oxygen, is estimated by $|\dot{m}|\cdot\tau$ with
$\tau=112\,$s the sum of the 3 lifetimes. At the beginning, the
neutral cloud is idealized as spherical; when moving out after
the ionization, it is enlarged proportionally to heliocentric distance
in the two directions perpendicular to the radial direction.
In the radial direction, there is an enlargement due to thermal
velocity dispersion (which cannot take place in the two other
directions as the ions are bound to the magnetic field
lines by gyration).
Hence the cloud results as a rotationally symmetric ellipsoid at 1 AU.
Let $a$ be the semi-minor axis in the two directions perpendicular
to the radial direction and $b$ the semi-major axis in the
radial direction. Hence $a$ is $\approx 262\,$km$\cdot
1\,$AU$/2.19R_{\text{sun}}=2.6\cdot 10^7\,$m.
The kinetic temperature of oxygen ions in the solar wind is
$\approx 2 \cdot 10^6\,$K (Hefti et al. 1998), implying a thermal
velocity of $v_{\text{th}} \approx 50\,$km$\,$s$^{-1}$. This velocity distribution
is dominated by damped alfven waves. The correlation length of the velocity
fluctuations and its dependence on the distance from the sun
is not well constrained in the inner heliosphere. From the available
measurements reviewed in Tu and Marsch (1995) we estimate a correlation
length of 2$\times$ 10$^9$ m. We approximate the elongation $b$ of the
ellipsoid by a random walk with the step size being the linear expansion
of the ions while traveling one correlation length and the number of
steps being the number of correlation lengths in 1 AU:
\begin{equation}
b = v_{th} \frac{\lambda}{v_s} \sqrt{\frac{\Delta}{\lambda}}
  = \frac{v_{th}}{v_s} \sqrt{\lambda\Delta}
\end{equation}
Here $v_s$ is the solar wind velocity, $\lambda$ the correlation
length, and $\Delta$ = 1 AU the distance of the observer from the sun.
For $v_s$ = 400 km/s, $v_{th}$ = 50 km/s, and $\lambda$ = 2$\times$
10$^9$ m, \textit{b} is 2.2$\times$ 10$^9$ m.

%From the solar wind
%velocity profile used here follows that an O$^+$ ions needs
%$5.5$ days to move from the ascending node to $1\,$AU. Therefore,
%$b$ is $\approx 34R_{\text{sun}}$.

%
\subsection{Fluxes, duration, and probability for a detection by a spacecraft
at 1\,AU}

We now try to estimate if an ion detector on a spacecraft may be able to 
find ions originating from a sungrazing comet. We determine the oxygen flux at 
1\,AU and the probability that a spacecraft located at 1\,AU in the ecliptic 
plane passes through a cloud of material evaporated from a sungrazing comet. We
assume that the spacecraft is close to earth and at rest relative to earth.

The spacecraft crosses the ellipsoid in the direction of $a$ with the orbital 
velocity of the earth,
hence within $2a/30\,$km$\,$s$^{-1}\approx 29$ minutes. In
contrast, the ellipsoid needs $2b/400\,$km$\,$s$^{-1}= 3$ hours
to cross the line of $1\,$AU heliocentric distance. Therefore,
the spacecraft is able to observe oxygen ions from a sungrazer during
$\approx 29$ minutes.

The longitude $\Omega$ of the ascending node directly indicates
the position the spacecraft must have to be in the purely radial path
of the ions. The average of $\Omega$ yields the date 22nd September
and the standard deviation the period 3rd September to 11th October.
The velocity $v_r$ of the ions relative to the spacecraft is
$\sqrt{400^2+30^2}\approx 400\,$km$\,$s$^{-1}$.
The average mass flux $F$ (mass per area and time)
of all oxygen ions at the spacecraft is the product of the mass density in
the ellipsoid and $v_r$.
\begin{equation}
\label{eq:F}
F=\frac{16/18\cdot|\dot{m}|\cdot\tau}{\frac{4\pi}{3}a^2 b}\cdot v_r=
6.53\cdot10^{-18}\cdot|\dot{m}|
\end{equation}
Evaluating Eq. \ref{eq:mdot} at the ascending node gives
\begin{equation}
\label{eq:139}
|\dot{m}|=13.9\cdot(1-A)\cdot R^2
\end{equation}
$R$ is the radius at the ascending node and can be calculated
with Eq. \ref{eq:dRdtheta}:
$R=R_0+\int_{-\pi}^{-\omega}\frac{dR}{d\vartheta}d\vartheta
=R_0-9290\cdot(1-A)/\varrho$. Inserting this in Eq. \ref{eq:139} and
then Eq. \ref{eq:139} in Eq. \ref{eq:F} yields:
$$
\boxed{
F(R_0,\varrho,A)=
9.1\cdot 10^{-17}\cdot(1-A)\cdot\left(R_0-9290\cdot\frac{1-A}{\varrho}\right)^2}
$$

For a nucleus with albedo zero, a density of
$600\,$kg$\,$m$^{-3}$ and an initial radius just small enough
to be vaporized completely at perihelion,
$R_0=1/2\cdot\Delta R(A=0,\varrho=600\,$kg$\,$m$^{-3},
q=1.307R_{\text{sun}})\approx 28\,$m,
Table I shows the oxygen ion fluxes, in particles
per m$^2$ and s, at 1 AU for each charge state. The expected cometary
fluxes of $\approx 10^{11}$ m$^{-3}$ s$^{-1}$ are approximately two orders
of magnitude above the oxygen flux in the solar wind at 1 AU.
Especially O$^{4+}$, which is virtually missing in the solar wind, could be
detected easily. On the other hand, we note that the initial radius of the
comet must be above 15\,m. A smaller comet sublimates completely before
reaching the ascending node.

From orbital geometry, the position the spacecraft must have
to measure sungrazer oxygen ions is fixed, and hence also
the approximative period during the year: 3rd September till
11th October. Therefore, a detection is a question of comet orbit timing:
what is the probability
that a sungrazer passes the ascending node in a period such
that its oxygen ions can meet the spacecraft? Between 1996 and June 2001,
about 300 sungrazers have been discovered. Some of them might be too small to
reach the ascending node and therefore to be detected. On the other hand,
SOHO might not have detected all sungrazers. For a rough estimate of the
detection probability, we assume that there were 300 detectable sungrazers
in the 5.5 years SOHO has been in operation.

Each sungrazer produces an ellipsoid of ions that needs
$\approx 3$ hours to pass the line of $1\,$AU
heliocentric distance. If there were always at least $3$ hours
between the ascending node passage of two sungrazers --
an assumption which is likely to be fulfilled --, then
there would be $\approx 37.5$ days per 5.5 years ($\approx$7 days per year) when
a spacecraft would pass sungrazer oxygen ions around the 22nd September.
This yields an estimation for the
measurement probability: $\approx 1.9\,\%$ per year,
restricted to the approximative period 3rd September till
11th October.

While the sublimation flux from a sungrazing comet at a heliocentric
distance of 2 solar radii could be easily measured at 1 AU, a spacecraft located
at 1\,AU will pass through a cloud of material from a sungrazing 
comet of the Kreutz family only approximately once in 50 years.

% ******************************************************************
\section{Conclusions}
Using an H$_2$O sublimation model for cometary nuclei, we have
derived upper size limits for sungrazers which are completely destroyed during
perihelion passage. Considering sublimation only, the  size limits can be 
expressed by the following analytical approximation to our numerical model:
$$
R_0\leq\frac{P_{\text{sun}}\cdot 0.85}{8\cdot\sqrt{2GM_{\text{sun}}}}\cdot\frac{1-A}{\varrho L\sqrt{q}}\approx 10^9\cdot\frac{1-A}{\varrho\sqrt{q}}
$$
By setting the albedo to zero and the perihelion distance
to one solar radius, one gets a general restriction valid
for all sungrazers completely destroyed by sublimation alone:
$$
R_0\leq 3.8\cdot 10^4\cdot\varrho^{-1}
$$
For the density of porous ice, $\approx 600\,$kg$\,$m$^{-3}$, which
was derived for comet Shoemaker-Levy 9 in Asphaug and Benz 1996,
the upper limit in size is $\approx 63\,$m. 

Depending on the spatial distribution of the sublimation and the density of the
cometary ice, the upper size limit may be increased by tidal disruption of the 
comet. Sublimation pressure will inhibit tidal disruption of an
isotropically sublimating comet. In the case of completely anisotropic
sublimation (no activity on the night side of the comet), the size limit increases
to $\approx 100\,$ m (at $600\,$ kg$\,$m$^{-3}$) 

We estimated a probability of $\approx 2\,\%$ per year,
restricted to the approximative period around the 22nd September
from 3rd September till 11th October, of a spacecraft detection of  oxygen ions
produced in the sublimation of a sungrazer. The duration
of the passage of the spacecraft through the ion stream is $\approx 29$ 
minutes. For the average sungrazer orbit, the charge state distribution
of the oxygen ions is given in Table I. The total mass flux $F$
for all charge states, in kg per m$^2$ and s,
depends on the cometary properties $R_0$, $\varrho$ and $A$ and
is estimated by
$$
F(R_0,\varrho,A)=
9.1\cdot 10^{-19}\cdot(1-A)\cdot\left(R_0-9290\cdot\frac{1-A}{\varrho}\right)^2
$$
For $R_0=28\,$m, $\varrho=600\,$kg$\,$m$^{-3}$ and $A=0$,
the fluxes for the different charge states are given in Table I, in particles 
per m$^2$ and s, and are well above detection
threshold. Unfortunately, the probability is quite low that a spacecraft 
actually passed through an ion cloud from a sublimated comet.
\iceskip
Comparing the absolute magnitudes of large sungrazers like
Ikeya-Seki and Pereyra with those of the
SOHO sungrazers shows a difference of about 15 magnitudes.
This implies that the large sungrazers were $100^3=10^6$ times brighter.
If this brightness can be scaled with the cometary surface,
then Ikeya-Seki and Pereyra were approximately 1000 times
larger than the SOHO sungrazers. This is consistent with
a size for Ikeya-Seki and Pereyra of a few kilometers
and a few meters for the SOHO sungrazers. Limiting the sizes
of Ikeya-Seki and Pereyra to 10$\,$km, and assuming this
estimation is good within one order of magnitude, the brightness
comparison provides additional evidence for a size of the SOHO sungrazers
of less than 100$\,$m.

If the SOHO sungrazers are fragments of a disruption several
revolutions ago, they sublimated continuously by $\Delta R$
per orbit. In this case it seems unlikely, that all fragments
disappear during the same orbit, as we observe now. Therefore,
we conclude that the SOHO sungrazers are remnants of a disruption
that occurred during the last perihelion passage,
approximately 1000 or 2000 years ago.  The huge number of
remnants indicates that the parent body was highly fragile, presumably
a rubble pile. The wide spread of the SOHO sungrazers along their orbit
shows that the remnants are not gravitationally bound. Hence
there must have been a mechanism that allowed the fragments
to overcome the gravitational binding of their parent body. The acceleration 
due to the sublimation pressure is different for different
fragments: The force varies with $R^2$, whereas the mass
varies with $R^3$. Hence even two bodies of equal shape
but different size will be accelerated differently and move
away from each other, unless the sublimation pressure is too low
to overcome gravitational binding. At this point, we cannot answer
this question, but one destruction mechanism that explains
the wide spread along the orbit, is tidal disruption. This would
also explain why we see a more or less constant flux of
sungrazers. A tidal disruption causes a wide spread of the
cometesimals. On the other hand, it is unlikely that a nearly
constant flux of 300 fragments can be produced
by a thermal stress splitting.

Since tidal forces are too weak to disrupt any material of appreciable
strength, the fragments of the tidal disruption of a comet cannot be
smaller than the cometesimals of that comet. If the SOHO sungrazers were
created by tidal disruption, their upper size limits
are also upper limits for the size of the cometesimals of their parent
body.

While the SOHO sungrazers would in this case consist of
either a few or only one cometesimal, larger remnants still
consist of many cometesimals and in some cases can
be disrupted further by tidal forces. For example, comet Ikeya-Seki broke
into three pieces when approaching perihelion. This is consistent with the
scenario for the disruption of a rubble pile comet on Ikeya-Seki's
orbit. Numerical solution of Eq. \ref{bf} predicts a mass of the
largest remnant of about half the parent body mass for a tidal disruption
on Ikeya-Seki's orbit ($q$ = 1.68 R$_{sun}$). The condition for tidal breakup
that tidal stress must be larger than sublimation pressure yields a lower
limit for the size of Ikeya-Seki in the case of isotropic sublimation. 
Replacing R$_{sun}$ in eq.\,\ref{eq:disr1} by the perihelion distance of 
Ikeya-Seki and assuming $\varrho$ = 600 kg m$^{-3}$ results in a radius of 
comet Ikeya-Seki of at least 3.5\,km. We note that this limit is valid in the
case of isotropic sublimation only.

The sungrazing state of comets is short-lived. A large sungrazer like the
progenitor of the Kreutz-family disrupts into several large fragments with a 
size of perhaps a few kms and many small ones like the sungrazers detected 
by SOHO. While the small fragments do not survive the next perihelion passage, 
the large ones will be destroyed after several tens of orbits (corresponding to
several 10$^4$ years for the Kreutz comets) if destructed by
sublimation alone. If further disruption occurs (as in the case of comet
Ikeya-Seki), the lifetime is even shorter. For this reason there are not many
sungrazing comets visible at a given time although the sungrazing state may
be a frequent cometary end state (Bailey et al 1992).

\section*{Acknowledgements}
The authors would like to thank Douglas Biesecker for
providing preliminary sungrazer orbit elements. Robert
Wimmer-Schweingruber was helpful to calculate the oxygen
ion fluxes. Eckart Marsch gave valuable input to the calculation of the
expansion of a cloud of oxygen ions in the solar wind. We acknowledge valuable 
discussions with Karine Issautier and Frederic Buclin. This work was supported 
in part by the Swiss National Science Foundation.

%\end{appendix}

%
\end{document}

%% file: fig1.eepic
\setlength{\unitlength}{0.00087489in}
\begingroup\makeatletter\ifx\SetFigFont\undefined%
\gdef\SetFigFont#1#2#3#4#5{%
  \reset@font\fontsize{#1}{#2pt}%
  \fontfamily{#3}\fontseries{#4}\fontshape{#5}%
  \selectfont}%
\fi\endgroup%
{\renewcommand{\dashlinestretch}{30}
\begin{picture}(3290,879)(0,-10)
\path(75,196)(75,646)(525,646)(525,196)
\path(525,646)(975,646)(975,196)
\path(975,646)(1425,646)(1425,196)
\path(75,196)(3000,196)
\blacken\path(2880.000,166.000)(3000.000,196.000)(2880.000,226.000)(2880.000,166.000)
\path(2325,196)(2325,646)(2775,646)(2775,196)
\put(869,1){\makebox(0,0)[lb]{\smash{{{\SetFigFont{10}{12.0}{\familydefault}{\mddefault}{\updefault}$R_2$}}}}}
\put(16,0){\makebox(0,0)[lb]{\smash{{{\SetFigFont{10}{12.0}{\familydefault}{\mddefault}{\updefault}$0$}}}}}
\put(419,1){\makebox(0,0)[lb]{\smash{{{\SetFigFont{10}{12.0}{\familydefault}{\mddefault}{\updefault}$R_1$}}}}}
\put(0,728){\makebox(0,0)[lb]{\smash{{{\SetFigFont{10}{12.0}{\familydefault}{\mddefault}{\updefault}$q_0$}}}}}
\put(1349,729){\makebox(0,0)[lb]{\smash{{{\SetFigFont{10}{12.0}{\familydefault}{\mddefault}{\updefault}$q_3$}}}}}
\put(900,729){\makebox(0,0)[lb]{\smash{{{\SetFigFont{10}{12.0}{\familydefault}{\mddefault}{\updefault}$q_2$}}}}}
\put(450,729){\makebox(0,0)[lb]{\smash{{{\SetFigFont{10}{12.0}{\familydefault}{\mddefault}{\updefault}$q_1$}}}}}
\put(2136,730){\makebox(0,0)[lb]{\smash{{{\SetFigFont{10}{12.0}{\familydefault}{\mddefault}{\updefault}$q_{l-1}$}}}}}
\put(1320,2){\makebox(0,0)[lb]{\smash{{{\SetFigFont{10}{12.0}{\familydefault}{\mddefault}{\updefault}$R_3$}}}}}
\put(2135,2){\makebox(0,0)[lb]{\smash{{{\SetFigFont{10}{12.0}{\familydefault}{\mddefault}{\updefault}$R_{l-1}$}}}}}
\put(1703,729){\makebox(0,0)[lb]{\smash{{{\SetFigFont{10}{12.0}{\familydefault}{\mddefault}{\updefault}$\dots$}}}}}
\put(3075,150){\makebox(0,0)[lb]{\smash{{{\SetFigFont{10}{12.0}{\familydefault}{\mddefault}{\updefault}$r$}}}}}
\put(2677,1){\makebox(0,0)[lb]{\smash{{{\SetFigFont{10}{12.0}{\familydefault}{\mddefault}{\updefault}$R_l$}}}}}
\put(2692,729){\makebox(0,0)[lb]{\smash{{{\SetFigFont{10}{12.0}{\familydefault}{\mddefault}{\updefault}$q_l$}}}}}
\put(1725,2){\makebox(0,0)[lb]{\smash{{{\SetFigFont{10}{12.0}{\familydefault}{\mddefault}{\updefault}$\dots$}}}}}
\put(1781,370){\makebox(0,0)[lb]{\smash{{{\SetFigFont{10}{12.0}{\familydefault}{\mddefault}{\updefault}$\dots$}}}}}
\put(160,381){\makebox(0,0)[lb]{\smash{{{\SetFigFont{10}{12.0}{\familydefault}{\mddefault}{\updefault}$T_{1/2}$}}}}}
\put(612,374){\makebox(0,0)[lb]{\smash{{{\SetFigFont{10}{12.0}{\familydefault}{\mddefault}{\updefault}$T_{3/2}$}}}}}
\put(1057,369){\makebox(0,0)[lb]{\smash{{{\SetFigFont{10}{12.0}{\familydefault}{\mddefault}{\updefault}$T_{5/2}$}}}}}
\put(2340,373){\makebox(0,0)[lb]{\smash{{{\SetFigFont{10}{12.0}{\familydefault}{\mddefault}{\updefault}$T_{l-1/2}$}}}}}
\end{picture}
}

%% file: fig8.tex
% GNUPLOT: LaTeX picture
\setlength{\unitlength}{0.240900pt}
\ifx\plotpoint\undefined\newsavebox{\plotpoint}\fi
\sbox{\plotpoint}{\rule[-0.200pt]{0.400pt}{0.400pt}}%
\begin{picture}(900,540)(0,0)
\font\gnuplot=cmr10 at 11pt
\gnuplot
\sbox{\plotpoint}{\rule[-0.200pt]{0.400pt}{0.400pt}}%
\put(132.0,135.0){\rule[-0.200pt]{4.818pt}{0.400pt}}
\put(110,135){\makebox(0,0)[r]{0}}
\put(857.0,135.0){\rule[-0.200pt]{4.818pt}{0.400pt}}
\put(132.0,189.0){\rule[-0.200pt]{4.818pt}{0.400pt}}
\put(110,189){\makebox(0,0)[r]{0.2}}
\put(857.0,189.0){\rule[-0.200pt]{4.818pt}{0.400pt}}
\put(132.0,243.0){\rule[-0.200pt]{4.818pt}{0.400pt}}
\put(110,243){\makebox(0,0)[r]{0.4}}
\put(857.0,243.0){\rule[-0.200pt]{4.818pt}{0.400pt}}
\put(132.0,297.0){\rule[-0.200pt]{4.818pt}{0.400pt}}
\put(110,297){\makebox(0,0)[r]{0.6}}
\put(857.0,297.0){\rule[-0.200pt]{4.818pt}{0.400pt}}
\put(132.0,351.0){\rule[-0.200pt]{4.818pt}{0.400pt}}
\put(110,351){\makebox(0,0)[r]{0.8}}
\put(857.0,351.0){\rule[-0.200pt]{4.818pt}{0.400pt}}
\put(132.0,405.0){\rule[-0.200pt]{4.818pt}{0.400pt}}
\put(110,405){\makebox(0,0)[r]{1}}
\put(857.0,405.0){\rule[-0.200pt]{4.818pt}{0.400pt}}
\put(149.0,135.0){\rule[-0.200pt]{0.400pt}{4.818pt}}
\put(149,90){\makebox(0,0){-3}}
\put(149.0,385.0){\rule[-0.200pt]{0.400pt}{4.818pt}}
\put(267.0,135.0){\rule[-0.200pt]{0.400pt}{4.818pt}}
\put(267,90){\makebox(0,0){-2}}
\put(267.0,385.0){\rule[-0.200pt]{0.400pt}{4.818pt}}
\put(386.0,135.0){\rule[-0.200pt]{0.400pt}{4.818pt}}
\put(386,90){\makebox(0,0){-1}}
\put(386.0,385.0){\rule[-0.200pt]{0.400pt}{4.818pt}}
\put(505.0,135.0){\rule[-0.200pt]{0.400pt}{4.818pt}}
\put(505,90){\makebox(0,0){0}}
\put(505.0,385.0){\rule[-0.200pt]{0.400pt}{4.818pt}}
\put(623.0,135.0){\rule[-0.200pt]{0.400pt}{4.818pt}}
\put(623,90){\makebox(0,0){1}}
\put(623.0,385.0){\rule[-0.200pt]{0.400pt}{4.818pt}}
\put(742.0,135.0){\rule[-0.200pt]{0.400pt}{4.818pt}}
\put(742,90){\makebox(0,0){2}}
\put(742.0,385.0){\rule[-0.200pt]{0.400pt}{4.818pt}}
\put(860.0,135.0){\rule[-0.200pt]{0.400pt}{4.818pt}}
\put(860,90){\makebox(0,0){3}}
\put(860.0,385.0){\rule[-0.200pt]{0.400pt}{4.818pt}}
\put(132.0,135.0){\rule[-0.200pt]{179.470pt}{0.400pt}}
\put(877.0,135.0){\rule[-0.200pt]{0.400pt}{65.043pt}}
\put(132.0,405.0){\rule[-0.200pt]{179.470pt}{0.400pt}}
\put(504,23){\makebox(0,0){true anomaly $\vartheta$ [radians]}}
\put(504,473){\makebox(0,0){Sublimation/solar input}}
\put(132.0,135.0){\rule[-0.200pt]{0.400pt}{65.043pt}}
\sbox{\plotpoint}{\rule[-0.400pt]{0.800pt}{0.800pt}}%
\put(139,135){\usebox{\plotpoint}}
\put(140.84,135){\rule{0.800pt}{0.482pt}}
\multiput(140.34,135.00)(1.000,1.000){2}{\rule{0.800pt}{0.241pt}}
\put(141.84,137){\rule{0.800pt}{4.095pt}}
\multiput(141.34,137.00)(1.000,8.500){2}{\rule{0.800pt}{2.048pt}}
\put(142.84,154){\rule{0.800pt}{9.636pt}}
\multiput(142.34,154.00)(1.000,20.000){2}{\rule{0.800pt}{4.818pt}}
\put(143.84,194){\rule{0.800pt}{9.636pt}}
\multiput(143.34,194.00)(1.000,20.000){2}{\rule{0.800pt}{4.818pt}}
\put(144.84,234){\rule{0.800pt}{7.950pt}}
\multiput(144.34,234.00)(1.000,16.500){2}{\rule{0.800pt}{3.975pt}}
\put(145.84,267){\rule{0.800pt}{6.023pt}}
\multiput(145.34,267.00)(1.000,12.500){2}{\rule{0.800pt}{3.011pt}}
\put(146.84,292){\rule{0.800pt}{4.818pt}}
\multiput(146.34,292.00)(1.000,10.000){2}{\rule{0.800pt}{2.409pt}}
\put(147.84,312){\rule{0.800pt}{3.614pt}}
\multiput(147.34,312.00)(1.000,7.500){2}{\rule{0.800pt}{1.807pt}}
\put(148.84,327){\rule{0.800pt}{2.891pt}}
\multiput(148.34,327.00)(1.000,6.000){2}{\rule{0.800pt}{1.445pt}}
\put(149.84,339){\rule{0.800pt}{2.409pt}}
\multiput(149.34,339.00)(1.000,5.000){2}{\rule{0.800pt}{1.204pt}}
\put(150.84,349){\rule{0.800pt}{1.686pt}}
\multiput(150.34,349.00)(1.000,3.500){2}{\rule{0.800pt}{0.843pt}}
\put(151.84,356){\rule{0.800pt}{1.445pt}}
\multiput(151.34,356.00)(1.000,3.000){2}{\rule{0.800pt}{0.723pt}}
\put(152.84,362){\rule{0.800pt}{1.204pt}}
\multiput(152.34,362.00)(1.000,2.500){2}{\rule{0.800pt}{0.602pt}}
\put(153.84,367){\rule{0.800pt}{0.964pt}}
\multiput(153.34,367.00)(1.000,2.000){2}{\rule{0.800pt}{0.482pt}}
\put(154.84,371){\rule{0.800pt}{0.723pt}}
\multiput(154.34,371.00)(1.000,1.500){2}{\rule{0.800pt}{0.361pt}}
\put(155.84,374){\rule{0.800pt}{0.723pt}}
\multiput(155.34,374.00)(1.000,1.500){2}{\rule{0.800pt}{0.361pt}}
\put(156.84,377){\rule{0.800pt}{0.482pt}}
\multiput(156.34,377.00)(1.000,1.000){2}{\rule{0.800pt}{0.241pt}}
\put(158.34,379){\rule{0.800pt}{0.723pt}}
\multiput(157.34,379.00)(2.000,1.500){2}{\rule{0.800pt}{0.361pt}}
\put(161,380.84){\rule{0.482pt}{0.800pt}}
\multiput(161.00,380.34)(1.000,1.000){2}{\rule{0.241pt}{0.800pt}}
\put(163,381.84){\rule{0.241pt}{0.800pt}}
\multiput(163.00,381.34)(0.500,1.000){2}{\rule{0.120pt}{0.800pt}}
\put(164,382.84){\rule{0.241pt}{0.800pt}}
\multiput(164.00,382.34)(0.500,1.000){2}{\rule{0.120pt}{0.800pt}}
\put(139.0,135.0){\usebox{\plotpoint}}
\put(166,383.84){\rule{0.241pt}{0.800pt}}
\multiput(166.00,383.34)(0.500,1.000){2}{\rule{0.120pt}{0.800pt}}
\put(165.0,385.0){\usebox{\plotpoint}}
\put(171,384.84){\rule{0.241pt}{0.800pt}}
\multiput(171.00,384.34)(0.500,1.000){2}{\rule{0.120pt}{0.800pt}}
\put(172,384.84){\rule{0.482pt}{0.800pt}}
\multiput(172.00,385.34)(1.000,-1.000){2}{\rule{0.241pt}{0.800pt}}
\put(167.0,386.0){\rule[-0.400pt]{0.964pt}{0.800pt}}
\put(180,383.84){\rule{0.241pt}{0.800pt}}
\multiput(180.00,384.34)(0.500,-1.000){2}{\rule{0.120pt}{0.800pt}}
\put(174.0,386.0){\rule[-0.400pt]{1.445pt}{0.800pt}}
\put(184,382.84){\rule{0.482pt}{0.800pt}}
\multiput(184.00,383.34)(1.000,-1.000){2}{\rule{0.241pt}{0.800pt}}
\put(181.0,385.0){\usebox{\plotpoint}}
\put(189,381.84){\rule{0.241pt}{0.800pt}}
\multiput(189.00,382.34)(0.500,-1.000){2}{\rule{0.120pt}{0.800pt}}
\put(186.0,384.0){\usebox{\plotpoint}}
\put(194,380.84){\rule{0.241pt}{0.800pt}}
\multiput(194.00,381.34)(0.500,-1.000){2}{\rule{0.120pt}{0.800pt}}
\put(190.0,383.0){\rule[-0.400pt]{0.964pt}{0.800pt}}
\put(198,379.84){\rule{0.241pt}{0.800pt}}
\multiput(198.00,380.34)(0.500,-1.000){2}{\rule{0.120pt}{0.800pt}}
\put(195.0,382.0){\usebox{\plotpoint}}
\put(203,378.84){\rule{0.241pt}{0.800pt}}
\multiput(203.00,379.34)(0.500,-1.000){2}{\rule{0.120pt}{0.800pt}}
\put(199.0,381.0){\rule[-0.400pt]{0.964pt}{0.800pt}}
\put(208,377.84){\rule{0.241pt}{0.800pt}}
\multiput(208.00,378.34)(0.500,-1.000){2}{\rule{0.120pt}{0.800pt}}
\put(204.0,380.0){\rule[-0.400pt]{0.964pt}{0.800pt}}
\put(214,376.84){\rule{0.241pt}{0.800pt}}
\multiput(214.00,377.34)(0.500,-1.000){2}{\rule{0.120pt}{0.800pt}}
\put(209.0,379.0){\rule[-0.400pt]{1.204pt}{0.800pt}}
\put(220,375.84){\rule{0.482pt}{0.800pt}}
\multiput(220.00,376.34)(1.000,-1.000){2}{\rule{0.241pt}{0.800pt}}
\put(215.0,378.0){\rule[-0.400pt]{1.204pt}{0.800pt}}
\put(227,374.84){\rule{0.241pt}{0.800pt}}
\multiput(227.00,375.34)(0.500,-1.000){2}{\rule{0.120pt}{0.800pt}}
\put(222.0,377.0){\rule[-0.400pt]{1.204pt}{0.800pt}}
\put(235,373.84){\rule{0.241pt}{0.800pt}}
\multiput(235.00,374.34)(0.500,-1.000){2}{\rule{0.120pt}{0.800pt}}
\put(228.0,376.0){\rule[-0.400pt]{1.686pt}{0.800pt}}
\put(242,372.84){\rule{0.241pt}{0.800pt}}
\multiput(242.00,373.34)(0.500,-1.000){2}{\rule{0.120pt}{0.800pt}}
\put(236.0,375.0){\rule[-0.400pt]{1.445pt}{0.800pt}}
\put(252,371.84){\rule{0.241pt}{0.800pt}}
\multiput(252.00,372.34)(0.500,-1.000){2}{\rule{0.120pt}{0.800pt}}
\put(243.0,374.0){\rule[-0.400pt]{2.168pt}{0.800pt}}
\put(261,370.84){\rule{0.241pt}{0.800pt}}
\multiput(261.00,371.34)(0.500,-1.000){2}{\rule{0.120pt}{0.800pt}}
\put(253.0,373.0){\rule[-0.400pt]{1.927pt}{0.800pt}}
\put(271,369.84){\rule{0.241pt}{0.800pt}}
\multiput(271.00,370.34)(0.500,-1.000){2}{\rule{0.120pt}{0.800pt}}
\put(262.0,372.0){\rule[-0.400pt]{2.168pt}{0.800pt}}
\put(284,368.84){\rule{0.241pt}{0.800pt}}
\multiput(284.00,369.34)(0.500,-1.000){2}{\rule{0.120pt}{0.800pt}}
\put(272.0,371.0){\rule[-0.400pt]{2.891pt}{0.800pt}}
\put(297,367.84){\rule{0.241pt}{0.800pt}}
\multiput(297.00,368.34)(0.500,-1.000){2}{\rule{0.120pt}{0.800pt}}
\put(285.0,370.0){\rule[-0.400pt]{2.891pt}{0.800pt}}
\put(312,366.84){\rule{0.241pt}{0.800pt}}
\multiput(312.00,367.34)(0.500,-1.000){2}{\rule{0.120pt}{0.800pt}}
\put(298.0,369.0){\rule[-0.400pt]{3.373pt}{0.800pt}}
\put(328,365.84){\rule{0.241pt}{0.800pt}}
\multiput(328.00,366.34)(0.500,-1.000){2}{\rule{0.120pt}{0.800pt}}
\put(313.0,368.0){\rule[-0.400pt]{3.613pt}{0.800pt}}
\put(347,364.84){\rule{0.241pt}{0.800pt}}
\multiput(347.00,365.34)(0.500,-1.000){2}{\rule{0.120pt}{0.800pt}}
\put(329.0,367.0){\rule[-0.400pt]{4.336pt}{0.800pt}}
\put(370,363.84){\rule{0.241pt}{0.800pt}}
\multiput(370.00,364.34)(0.500,-1.000){2}{\rule{0.120pt}{0.800pt}}
\put(348.0,366.0){\rule[-0.400pt]{5.300pt}{0.800pt}}
\put(398,362.84){\rule{0.241pt}{0.800pt}}
\multiput(398.00,363.34)(0.500,-1.000){2}{\rule{0.120pt}{0.800pt}}
\put(371.0,365.0){\rule[-0.400pt]{6.504pt}{0.800pt}}
\put(439,361.84){\rule{0.241pt}{0.800pt}}
\multiput(439.00,362.34)(0.500,-1.000){2}{\rule{0.120pt}{0.800pt}}
\put(399.0,364.0){\rule[-0.400pt]{9.636pt}{0.800pt}}
\put(561,361.84){\rule{0.241pt}{0.800pt}}
\multiput(561.00,361.34)(0.500,1.000){2}{\rule{0.120pt}{0.800pt}}
\put(440.0,363.0){\rule[-0.400pt]{29.149pt}{0.800pt}}
\put(602,362.84){\rule{0.482pt}{0.800pt}}
\multiput(602.00,362.34)(1.000,1.000){2}{\rule{0.241pt}{0.800pt}}
\put(562.0,364.0){\rule[-0.400pt]{9.636pt}{0.800pt}}
\put(631,363.84){\rule{0.482pt}{0.800pt}}
\multiput(631.00,363.34)(1.000,1.000){2}{\rule{0.241pt}{0.800pt}}
\put(604.0,365.0){\rule[-0.400pt]{6.504pt}{0.800pt}}
\put(654,364.84){\rule{0.241pt}{0.800pt}}
\multiput(654.00,364.34)(0.500,1.000){2}{\rule{0.120pt}{0.800pt}}
\put(633.0,366.0){\rule[-0.400pt]{5.059pt}{0.800pt}}
\put(674,365.84){\rule{0.241pt}{0.800pt}}
\multiput(674.00,365.34)(0.500,1.000){2}{\rule{0.120pt}{0.800pt}}
\put(655.0,367.0){\rule[-0.400pt]{4.577pt}{0.800pt}}
\put(692,366.84){\rule{0.241pt}{0.800pt}}
\multiput(692.00,366.34)(0.500,1.000){2}{\rule{0.120pt}{0.800pt}}
\put(675.0,368.0){\rule[-0.400pt]{4.095pt}{0.800pt}}
\put(707,367.84){\rule{0.241pt}{0.800pt}}
\multiput(707.00,367.34)(0.500,1.000){2}{\rule{0.120pt}{0.800pt}}
\put(693.0,369.0){\rule[-0.400pt]{3.373pt}{0.800pt}}
\put(722,368.84){\rule{0.241pt}{0.800pt}}
\multiput(722.00,368.34)(0.500,1.000){2}{\rule{0.120pt}{0.800pt}}
\put(708.0,370.0){\rule[-0.400pt]{3.373pt}{0.800pt}}
\put(735,369.84){\rule{0.241pt}{0.800pt}}
\multiput(735.00,369.34)(0.500,1.000){2}{\rule{0.120pt}{0.800pt}}
\put(723.0,371.0){\rule[-0.400pt]{2.891pt}{0.800pt}}
\put(746,370.84){\rule{0.241pt}{0.800pt}}
\multiput(746.00,370.34)(0.500,1.000){2}{\rule{0.120pt}{0.800pt}}
\put(736.0,372.0){\rule[-0.400pt]{2.409pt}{0.800pt}}
\put(758,371.84){\rule{0.241pt}{0.800pt}}
\multiput(758.00,371.34)(0.500,1.000){2}{\rule{0.120pt}{0.800pt}}
\put(747.0,373.0){\rule[-0.400pt]{2.650pt}{0.800pt}}
\put(769,372.84){\rule{0.241pt}{0.800pt}}
\multiput(769.00,372.34)(0.500,1.000){2}{\rule{0.120pt}{0.800pt}}
\put(759.0,374.0){\rule[-0.400pt]{2.409pt}{0.800pt}}
\put(778,373.84){\rule{0.482pt}{0.800pt}}
\multiput(778.00,373.34)(1.000,1.000){2}{\rule{0.241pt}{0.800pt}}
\put(770.0,375.0){\rule[-0.400pt]{1.927pt}{0.800pt}}
\put(788,374.84){\rule{0.241pt}{0.800pt}}
\multiput(788.00,374.34)(0.500,1.000){2}{\rule{0.120pt}{0.800pt}}
\put(780.0,376.0){\rule[-0.400pt]{1.927pt}{0.800pt}}
\put(797,375.84){\rule{0.241pt}{0.800pt}}
\multiput(797.00,375.34)(0.500,1.000){2}{\rule{0.120pt}{0.800pt}}
\put(789.0,377.0){\rule[-0.400pt]{1.927pt}{0.800pt}}
\put(806,376.84){\rule{0.241pt}{0.800pt}}
\multiput(806.00,376.34)(0.500,1.000){2}{\rule{0.120pt}{0.800pt}}
\put(798.0,378.0){\rule[-0.400pt]{1.927pt}{0.800pt}}
\put(815,377.84){\rule{0.241pt}{0.800pt}}
\multiput(815.00,377.34)(0.500,1.000){2}{\rule{0.120pt}{0.800pt}}
\put(807.0,379.0){\rule[-0.400pt]{1.927pt}{0.800pt}}
\put(826,378.84){\rule{0.241pt}{0.800pt}}
\multiput(826.00,378.34)(0.500,1.000){2}{\rule{0.120pt}{0.800pt}}
\put(816.0,380.0){\rule[-0.400pt]{2.409pt}{0.800pt}}
\put(834,378.84){\rule{0.241pt}{0.800pt}}
\multiput(834.00,379.34)(0.500,-1.000){2}{\rule{0.120pt}{0.800pt}}
\put(827.0,381.0){\rule[-0.400pt]{1.686pt}{0.800pt}}
\put(840,377.84){\rule{0.241pt}{0.800pt}}
\multiput(840.00,378.34)(0.500,-1.000){2}{\rule{0.120pt}{0.800pt}}
\put(835.0,380.0){\rule[-0.400pt]{1.204pt}{0.800pt}}
\put(843,376.84){\rule{0.241pt}{0.800pt}}
\multiput(843.00,377.34)(0.500,-1.000){2}{\rule{0.120pt}{0.800pt}}
\put(844,375.84){\rule{0.241pt}{0.800pt}}
\multiput(844.00,376.34)(0.500,-1.000){2}{\rule{0.120pt}{0.800pt}}
\put(845,374.84){\rule{0.241pt}{0.800pt}}
\multiput(845.00,375.34)(0.500,-1.000){2}{\rule{0.120pt}{0.800pt}}
\put(846,373.84){\rule{0.241pt}{0.800pt}}
\multiput(846.00,374.34)(0.500,-1.000){2}{\rule{0.120pt}{0.800pt}}
\put(847,372.84){\rule{0.241pt}{0.800pt}}
\multiput(847.00,373.34)(0.500,-1.000){2}{\rule{0.120pt}{0.800pt}}
\put(848,371.84){\rule{0.241pt}{0.800pt}}
\multiput(848.00,372.34)(0.500,-1.000){2}{\rule{0.120pt}{0.800pt}}
\put(847.84,371){\rule{0.800pt}{0.482pt}}
\multiput(847.34,372.00)(1.000,-1.000){2}{\rule{0.800pt}{0.241pt}}
\put(848.84,368){\rule{0.800pt}{0.723pt}}
\multiput(848.34,369.50)(1.000,-1.500){2}{\rule{0.800pt}{0.361pt}}
\put(849.84,366){\rule{0.800pt}{0.482pt}}
\multiput(849.34,367.00)(1.000,-1.000){2}{\rule{0.800pt}{0.241pt}}
\put(850.84,362){\rule{0.800pt}{0.964pt}}
\multiput(850.34,364.00)(1.000,-2.000){2}{\rule{0.800pt}{0.482pt}}
\put(851.84,358){\rule{0.800pt}{0.964pt}}
\multiput(851.34,360.00)(1.000,-2.000){2}{\rule{0.800pt}{0.482pt}}
\put(852.84,353){\rule{0.800pt}{1.204pt}}
\multiput(852.34,355.50)(1.000,-2.500){2}{\rule{0.800pt}{0.602pt}}
\put(853.84,347){\rule{0.800pt}{1.445pt}}
\multiput(853.34,350.00)(1.000,-3.000){2}{\rule{0.800pt}{0.723pt}}
\put(854.84,339){\rule{0.800pt}{1.927pt}}
\multiput(854.34,343.00)(1.000,-4.000){2}{\rule{0.800pt}{0.964pt}}
\put(855.84,329){\rule{0.800pt}{2.409pt}}
\multiput(855.34,334.00)(1.000,-5.000){2}{\rule{0.800pt}{1.204pt}}
\put(856.84,317){\rule{0.800pt}{2.891pt}}
\multiput(856.34,323.00)(1.000,-6.000){2}{\rule{0.800pt}{1.445pt}}
\put(857.84,301){\rule{0.800pt}{3.854pt}}
\multiput(857.34,309.00)(1.000,-8.000){2}{\rule{0.800pt}{1.927pt}}
\put(858.84,281){\rule{0.800pt}{4.818pt}}
\multiput(858.34,291.00)(1.000,-10.000){2}{\rule{0.800pt}{2.409pt}}
\put(859.84,256){\rule{0.800pt}{6.023pt}}
\multiput(859.34,268.50)(1.000,-12.500){2}{\rule{0.800pt}{3.011pt}}
\put(860.84,224){\rule{0.800pt}{7.709pt}}
\multiput(860.34,240.00)(1.000,-16.000){2}{\rule{0.800pt}{3.854pt}}
\put(861.84,187){\rule{0.800pt}{8.913pt}}
\multiput(861.34,205.50)(1.000,-18.500){2}{\rule{0.800pt}{4.457pt}}
\put(862.84,153){\rule{0.800pt}{8.191pt}}
\multiput(862.34,170.00)(1.000,-17.000){2}{\rule{0.800pt}{4.095pt}}
\put(863.84,137){\rule{0.800pt}{3.854pt}}
\multiput(863.34,145.00)(1.000,-8.000){2}{\rule{0.800pt}{1.927pt}}
\put(866,134.34){\rule{0.482pt}{0.800pt}}
\multiput(866.00,135.34)(1.000,-2.000){2}{\rule{0.241pt}{0.800pt}}
\put(841.0,379.0){\usebox{\plotpoint}}
\put(868.0,135.0){\usebox{\plotpoint}}
\end{picture}